\pdfoutput=1
\documentclass[nofootinbib,aps,prd,twocolumn,superscriptaddress]{revtex4-1}
\usepackage{color}
\usepackage{aas_macros}
\usepackage{natbib}
\usepackage{adjustbox}
\usepackage{todonotes}
\usepackage{hyperref}

\newcommand{\bea}{\begin{eqnarray}}
\newcommand{\be}{\begin{equation}}
\newcommand{\ben}{\begin{enumerate}}
\newcommand{\bi}{\begin{itemize}}
\newcommand{\eea}{\end{eqnarray}}
\newcommand{\ee}{\end{equation}}
\newcommand{\ei}{\end{itemize}}
\newcommand{\een}{\end{enumerate}}

\newcommand{\om}{\Omega_\mr m}

\newcommand{\sig}{\sigma_8}
\newcommand{\lcdm}{$\Lambda$CDM~}

\renewcommand{\d}{{\rm d}}
\newcommand{\pd}{P_{\delta}}

\newcommand{\vt}{\vartheta}

\newcommand{\mr}{\mathrm}
\newcommand{\pofk}{P_{\delta}(k,z)}

\newcommand{\imshape}{{\textsc{im3shape}}}
\newcommand{\ngmix}{\textsc{ngmix}}
\newcommand{\halofit}{{\textsc{halofit}}}

\defcitealias{heymans13}{H13}
\defcitealias{kilbinger13}{K13}
\defcitealias{Bo15}{Bo15}
\defcitealias{J15}{J15}
\defcitealias{Be15}{Be15}

\newcommand{\xip}{\xi_+}
\newcommand{\xim}{\xi_-}
\newcommand{\xipm}{\xi_\pm}

\newcommand{\CWR}[1]{} 

\begin{document}
\title[Cosmology from Cosmic Shear with DES SV Data]{Cosmology from Cosmic Shear with DES Science Verification Data}

\date{\today}


\author{The Dark Energy Survey Collaboration}
\noaffiliation
\author{T.~Abbott}
\affiliation{Cerro Tololo Inter-American Observatory, National Optical Astronomy Observatory, Casilla 603, La Serena, Chile}
\author{F.~B.~Abdalla}
\affiliation{Department of Physics \& Astronomy, University College London, Gower Street, London, WC1E 6BT, UK}
\author{S.~Allam}
\affiliation{Fermi National Accelerator Laboratory, P. O. Box 500, Batavia, IL 60510, USA}
\author{A.~Amara}
\affiliation{Department of Physics, ETH Zurich, Wolfgang-Pauli-Strasse 16, CH-8093 Zurich, Switzerland}
\author{J.~Annis}
\affiliation{Fermi National Accelerator Laboratory, P. O. Box 500, Batavia, IL 60510, USA}
\author{R.~Armstrong}
\affiliation{Department of Astrophysical Sciences, Princeton University, Peyton Hall, Princeton, NJ 08544, USA}
\author{D.~Bacon}
\affiliation{Institute of Cosmology \& Gravitation, University of Portsmouth, Portsmouth, PO1 3FX, UK}
\author{M.~Banerji}
\affiliation{Institute of Astronomy, University of Cambridge, Madingley Road, Cambridge CB3 0HA, UK}
\affiliation{Kavli Institute for Cosmology, University of Cambridge, Madingley Road, Cambridge CB3 0HA, UK}
\author{A.~H.~Bauer}
\affiliation{Institut de Ci\`encies de l'Espai, IEEC-CSIC, Campus UAB, Carrer de Can Magrans, s/n,  08193 Bellaterra, Barcelona, Spain}
% ...


\label{firstpage}
\begin{abstract}
We present the first constraints on cosmology from the Dark Energy Survey (DES), using weak lensing measurements from the preliminary Science Verification (SV) data.  We use 139 square degrees of SV data, which is less than 3\% of the full DES survey area.  Using cosmic shear 2-point measurements over three redshift bins we find $\sigma_8 (\Omega_{\rm m}/0.3)^{0.5} = 0.81 \pm 0.06$ (68\% confidence), after marginalising over 7 systematics parameters and 3 other cosmological parameters.  We examine the robustness of our results to the choice of data vector and systematics assumed, and find them to be stable. 
About $20$\% of our error bar comes from marginalising over shear and photometric redshift calibration uncertainties. 
The current state-of-the-art cosmic shear measurements from CFHTLenS are mildly discrepant with the cosmological constraints from Planck CMB data; our results are consistent with both datasets.
Our uncertainties are $\sim$30\% larger than those from CFHTLenS when we carry out a comparable analysis of the two datasets, which we attribute largely to the lower number density of our shear catalogue. 
We investigate constraints on dark energy and find that, with this small fraction of the full survey, the DES SV constraints make negligible impact on the Planck constraints.
The moderate disagreement between the CFHTLenS and Planck values of $\sigma_8 (\Omega_{\rm m}/0.3)^{0.5}$ is present regardless of the value of $w$.
\end{abstract}


\maketitle

\section{Introduction}\label{sec:intro}

The accelerated expansion of the Universe is the biggest mystery in modern cosmology. 
Many ongoing and future cosmology surveys are designed to shed new light on the potential causes of this acceleration using a range of techniques.
Many of these surveys will probe the acceleration using the subtle gravitational distortion of galaxy images, known as cosmic shear.
This method is particularly powerful because it is sensitive to both the expansion history of and the growth of structure in the Universe~\citep{detf,esoesa}. Measurement of both of these is important in trying to distinguish whether the acceleration is due to some substance in the Universe, dubbed dark energy, or whether General Relativity needs to be modified. 
Observations of cosmic shear offer the potential to elucidate the properties of dark energy and the nature of gravity. 
In addition, cosmic shear can constrain the amount and clustering of dark matter, which may help us to understand this mysterious constituent of the Universe and its role in galaxy formation.

Since the first detection of cosmic shear over a decade ago \cite{Bacon:2000yp,Kaiser:2000if,Wittman:2000tc,van_Waerbeke:2000rm}, a number of subsequent surveys led to steadily improved measurements \citep{Hoekstra02,hamana03,vanWaerbeke05,Jarvis06,Semboloni06,Massey07,Hetterscheidt07,Schrabback10}. 
More recently the Sloan Digital Sky Survey (SDSS) Stripe 82 region of 140 to 168 square degrees was analysed by \citet{Lin12} and \citet{Huff14}. The recent Deep Lens Survey (DLS) cosmological constraints by \citet{jee2013} used 20 square degrees of data taken with the Mosaic Imager on the Blanco telescope between 2000 and 2003. 
The Canada France Hawaii Telescope Lensing Survey (CFHTLenS, \cite{heymans12}) analysed $154$ square degrees of data taken as part of the Canada France Hawaii Telescope Legacy Survey (CFHTLS) between 2003 and 2009. 
CFHTLenS cosmology analyses included \citet{kilbinger13} (hereafter \citetalias{kilbinger13}), \citet{heymans13} (hereafter \citetalias{heymans13}), \citet{kitching14} and \citet{benjamin13}.
\citetalias{heymans13} performed  a six-redshift bin tomographic analysis, which is arguably the most constraining CFHTLenS result, since they marginalised over intrinsic alignments as well as cosmological parameters.
The Kilo-Degree Survey (KiDS) have just released a weak lensing analysis of $100$ square degrees of their survey and compare their cosmic shear measurements to predictions from CFHTLenS and Planck best-fit models \citep{Kuijken15}.

Cosmic shear measures the integrated fluctuations in matter density along a line of sight to the observed galaxies, with a weight kernel that peaks approximately half way to these galaxies.
This value can be compared with the clumpiness of the Universe at recombination observed in the temperature fluctuations of the Cosmic Microwave Background radiation (CMB), extrapolated to the present day using the parameters of $\Lambda$CDM derived from measurements of the CMB.
The most recent measurements from the Planck satellite \citep{planck2015} are in tension with CFHTLenS and some other low-redshift measurements, 
which could point to new physics such as non-negligible neutrino masses or a modified growth history~\citep{battye2014,beutler2014}. However,
as noted by \citet{maccrann2015}, massive neutrinos are not a natural explanation because they do not move the two sets of contours significantly closer together in the $\sigma_8$, $\Omega_{\rm m}$ plane.

Gravitational lensing of the Cosmic Microwave Background radiation provides additional information on the clumpiness of the low redshift Universe. It probes slightly higher redshifts than cosmic shear ($z\lesssim2$) and recent measurements have a constraining power comparable to that of current cosmic shear data~\citep{vanEngelen14,story14,planck2015lensing}. 

At present, three major ground-based cosmology surveys are in the process of taking high quality imaging data to measure cosmic shear: the KIlo-Degree Survey (KIDS)\footnote{\tt{http://kids.strw.leidenuniv.nl}} which uses the VLT Survey Telescope (VST), the Hyper Suprime-Cam (HSC) survey\footnote{\tt{http://www.naoj.org/Projects/HSC/HSCProject.html}} using the Subaru telescope, and the Dark Energy Survey (DES)\footnote{\tt{http://www.darkenergysurvey.org}} using the Blanco telescope. 
Furthermore, three new cosmology survey telescopes are under development for operation next decade, with designs tuned for cosmic shear measurements: the Large Synoptic Survey Telescope (LSST)\footnote{\tt{http://www.lsst.org}},  Euclid\footnote{\tt{http://sci.esa.int/euclid}} and the Wide Field InfraRed Survey Telescope (WFIRST)\footnote{\tt{http://wfirst.gsfc.nasa.gov}}.

Though one of the most cosmologically powerful techniques, cosmic shear is also among the most technically challenging. The lensing distortions are of order $2\%$, far smaller than the intrinsic ellipticities of typical galaxies. Therefore these distortions must be measured statistically, for example by averaging over an ensemble of galaxies within a patch of sky. To overcome statistical noise, millions of objects must be measured to high accuracy. The size and sky coverage of the next generation surveys will provide unprecedented statistical power. 

Before the power of these data can be exploited, however, a number of practical difficulties must be overcome. The most significant of these fall broadly into four categories. (i) Shape measurements must be carried out in the presence of noise, pixelisation, atmospheric distortion, and instrumental effects. These can be significantly larger than the shear signal itself. 
Even with perfect characterisation of these effects, biases can arise from e.g. imperfect knowledge of the intrinsic galaxy ellipticity distribution or morphology (see e.g. \citet{J15}).
(ii) To make useful cosmological inferences based on shear data one also needs accurate redshift information, but it is observationally infeasible to obtain spectroscopic redshifts for the large number of source galaxies. Instead one must rely on photometric redshift estimates (photo-\emph{z}s),
which are based on models of galaxy spectra, or spectroscopic training sets that may not be fully representative, and can therefore also suffer from biases (see e.g. \citet{Ma2006,BK07,Bernstein2009,MacDonald2010,Dahlen2013}). (iii) The cosmological lensing signal must be disentangled from intrinsic alignments (IAs). Systematic shape correlations can arise from tidal interactions between physically nearby galaxies during formation 
\cite{Djorgovsky1987,CKB01,CNP+01}.
Even excluding such pairs of objects, correlations between the intrinsic shapes of foreground galaxies and the shear of background galaxies can contaminate the cosmic shear signal. For recent reviews of the field see \citet{Kirk2015}, \citet{Joachimi2015} and \citet{Troxel20151}. 
(iv) The density fluctuations in the matter distribution must be predicted with sufficient precision to allow interpretation of the data. 
On small scales this is sensitive to uncertain effects of baryonic feedback on the underlying matter, which are not yet fully understood from hydrodynamic simulations. Ignoring these effects can induce significant bias in estimates of cosmological parameters \citep{vandalen11,sembolini11, harnois14}. For this reason cosmic shear studies commonly exclude the small scales where baryonic effects are expected to be strongest.

\begin{figure*}
\centerline{\includegraphics[width=14cm]{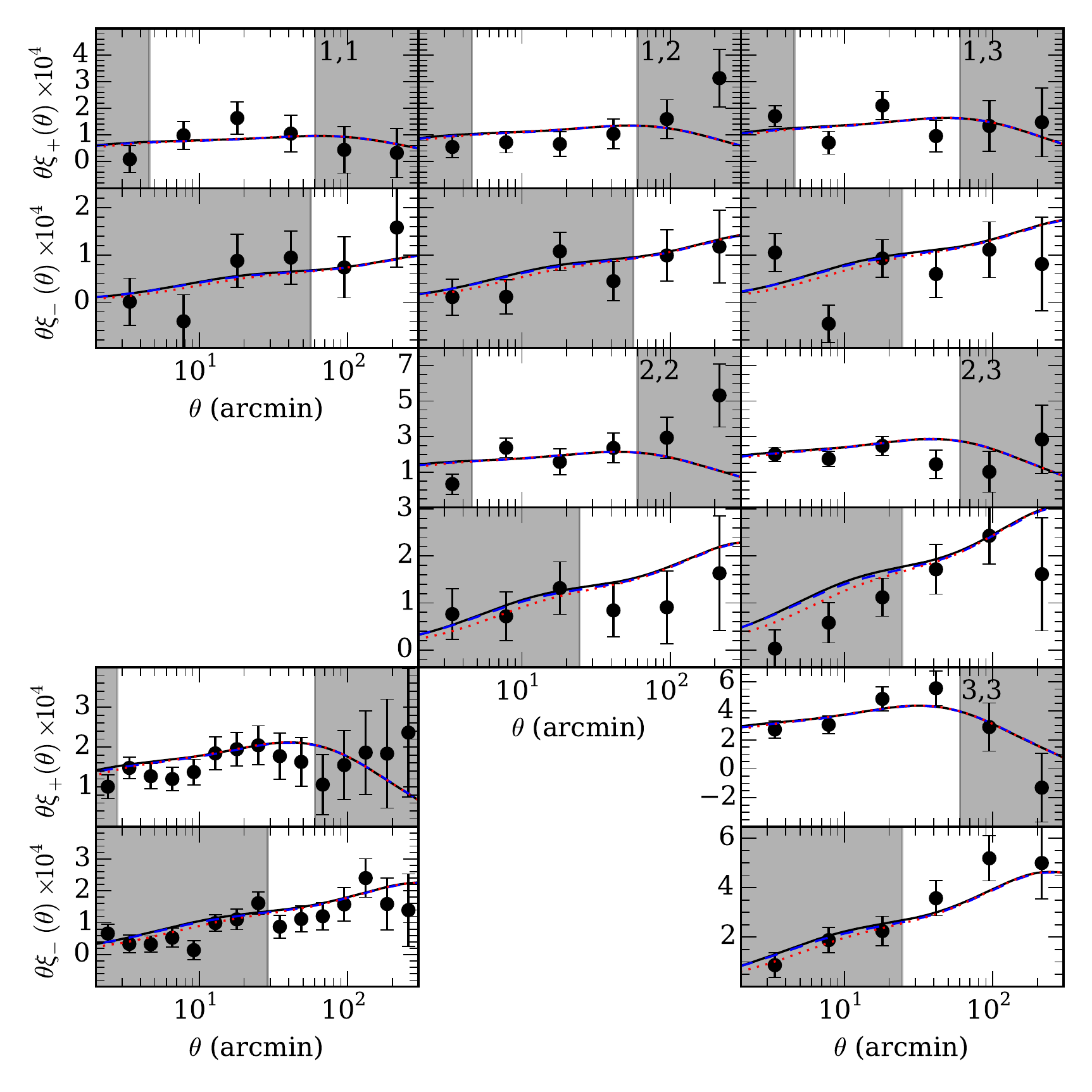}}
\caption{
DES SV shear two-point correlation function $\xipm$ measurements in each of the redshift bin pairings (from \citet{Be15}). The 3 redshift bins ranges are $0.3<z<0.55$, $0.55<z<0.83$ and $0.83<z<1.3$, and each galaxy is assigned to a redshift bin according to the mean of its photometric redshift probability distribution (or excluded if this value is outside the above ranges).
Alternating rows are $\xip$ and $\xim$, and the redshift bin combination is labelled in the upper right corner of each panel. The non-tomographic measurement is in the bottom left corner. The solid lines show the correlation functions computed for the best-fit Planck 2015 (TT + lowP) base \lcdm cosmology, using \halofit~\citep{smith03, takahashi2012} to model the non-linear matter power spectrum. The blue dashed lines (mostly obscured by the black lines) and red dotted lines assume the same cosmology but model nonlinear scales using FrankenEmu \citep{heitmann2014} (extended at high $k$ using the `CEp' presciption from \citet{harnois14}) and a prescription based on the OWLS `AGN' simulation \citep{schaye10} respectively. Points lying in grey regions are excluded from the analysis because they may be affected by either small-scale matter power spectrum uncertainty or large-scale additive shear bias, as explained in Section \ref{subsec:scales}.
}
\label{fig:xi_measurements}
\end{figure*}

In this paper we present the first cosmological constraints from the Dark Energy Survey, using the Science Verification data.
A detailed description of the methods and tests of galaxy shape measurements is given in \citet{J15} (hereafter \citetalias{J15}); the photometric redshift measurements are described in \citet{Bo15} (hereafter \citetalias{Bo15}) and the cosmic shear two-point function estimates and covariances are described in \citet{Be15} (hereafter \citetalias{Be15}). We focus here on cosmological constraints and their robustness to systematic effects and choice of data, as quantified in the companion papers. 
We describe the data in Section~\ref{sec:data} and present our main results in Section~\ref{sec:fid_constraints}. In Section~\ref{sec:data_vector} we discuss the impact of the choice of scales and two point statistic and we investigate the robustness of our main results to our assumptions about systematics in Section~\ref{sec:constraints}. 
Finally, we combine and compare our constraints with those from other surveys in Section~\ref{sec:combination} and conclude in Section~\ref{sec:conclusions}. More details on our intrinsic alignment models are given in Appendix \ref{sec:IA_appendix}.

\section{DES SV Data}\label{sec:data}

In this Section we overview some of the earlier work that provides essential ingredients for the cosmology analysis presented here.

\subsection{The survey}
\label{sec:survey}

The Dark Energy Survey (DES) is 
undertaking a five year programme of observations to image $\sim$5000 square degrees of the southern sky to $\sim24$th magnitude in the $grizY$ bands spanning 0.40-1.06~$\mu$m using the 570 megapixel imager DECam \citep{decam}. The survey will consist of $\sim$10 interlaced passes of 90~s exposures in each of $griz$ and 45~s in $Y$ over the full area.
The first weak lensing measurements from DES, using early commissioning data, were presented in \citet{melchior2014}.
Science Verification data were taken between November 2012 and February 2013, including a contiguous region in the South Pole Telescope East (SPTE) field, of which we use the $139$ square degrees presented in \citetalias{J15}.
A mass map of this field was presented in \citet{vikram15} and \citet{chang15}.  Significant improvements in instrument performance and image analysis techniques have been made during and since the Science Verification period, so that we can expect the DES lensing results to exceed those presented here in quality as well as quantity.

\subsection{Shear catalogues}

\label{sec:shapes}

The galaxy shape catalogue is discussed in detail in \citetalias{J15}, and is produced using two independent shear 
pipelines, \ngmix~\citep{sheldon14} and \imshape~\citep{zuntz2013}. Both shape measurement codes are based on model fitting techniques. Each object is fitted simultaneously to multiple reduced single epoch images. In addition to the intrinsic galaxy shape, the point spread function (PSF) and pixelisation are included in the model. The PSF is estimated separately on each exposure using the PSFEx package \citep{bertin}. The software measures the distortion kernel directly using bright stars. It then uses  polynomial interpolation across the image plane to estimate the PSF at specific galaxy locations. 
\citetalias{J15} carried out an extensive set of tests of the shear measurements and found them to be sufficiently free of systematics for the analysis presented here, provided that a small multiplicative uncertainty on the ellipticities is 
introduced. 

The raw number densities of the catalogues are $4.2$ and $6.9$ 
galaxies per square arcminute for \imshape~and \ngmix~respectively; weighted by signal-to-noise to get an effective number density we obtain 
3.7 and 5.7 
per square arcminute respectively\footnote{The definition of effective density used here differs from previous definitions in the literature; see \citetalias{J15}.}.
The fiducial catalogue 
is \ngmix; in Section~\ref{subsec:shear_cal} we show the results using \imshape~and the results ignoring the multiplicative bias uncertainty.

\subsubsection{Blinding}
To avoid experimenter bias the ellipticities that went into the 2-point functions used in this analysis were blinded by a constant scaling factor (between 0.9 and 1); this moved the contours in the $(\sigma_8$, $\Omega_m$) plane. Almost all adjustments to the analysis were completed before the blinding factor was removed, so any tendency to tune the results to match previous data or theory expectations was negated. After unblinding, some changes were made to the analysis: the maximum angular scale used for $\xip$ was changed from 30 to 60 arcmin as a result of an improvement in the additive systematics detailed in \citetalias{J15}. In particular, the shear difference correlation test in 8.6.2 of \citetalias{J15} significantly improved on large scales once a selection bias due to matching the two shear catalogs was accounted for. Additionally, a bug fix was applied to the weights in the \imshape~catalogue.

\subsection{Shear two-point function estimates}
\label{sec:shear2pt}

The first measurement of cosmic shear in DES SV is presented in \citetalias{Be15}. The primary two-point estimators used in that paper are the real-space angular 
shear correlation functions $\xipm$, defined as $\xipm(\theta)= \left \langle \gamma _{t}\gamma _{t} \right \rangle(\theta) \pm  \left \langle \gamma _{\times} \gamma _{\times} \right \rangle(\theta)$, where the angular brackets denote averaging over galaxy pairs separated by angle $\theta$ and $\gamma_{t,\times}$ are the tangential and cross shear components, measured relative to the separation vector.  
Our fiducial data vector,
the real-space angular correlation functions measured
in three tomographic bins, is
shown in Figure \ref{fig:xi_measurements}. The redshift bins used span: (1) $0.3<z<0.55$, (2) $0.55<z<0.83$, and (3) $0.83<z<1.30$.

\citetalias{Be15}
carry out a suite of systematics tests at the two-point level using $\xipm$ estimates and find the shear measurements suitable for the analysis described in this paper. They also calculate PolSpice \citep{szapudi2001} pseudo-$C_{\ell}$ estimates of the convergence power spectrum and Fourier band power estimates derived from linear combinations of $\xipm$ values \citep{becker2014}. In Section \ref{sec:choice_2pt} we compare cosmology constraints using our fiducial estimators, $\xipm$, to constraints using these.

\citetalias{Be15} estimate covariances of the two-point functions using both 126 simulated mock surveys and the halo model. The halo model covariance was computed from the {\sc CosmoLike} covariance module \citep{cosmolike}. It neglects the exact survey mask by assuming a simple symmetric geometry, but unlike the mock covariance it does not suffer from statistical uncertainties due to the estimation process. The 126 simulated mock surveys were generated from 21 large N-body simulations and hence include halo-sample variance, and the correct survey geometry. \citet{taylor2013} and \citet{DodelsonSchneider13} explore the implications on parameter constraints of noise in the covariance matrix estimate due to having a finite number of independent simulated surveys. The fiducial data vector used in this analysis has 36 data points, hence we can expect our reported parameter errorbars to be accurate to $\sim18\%$ (see \citetalias{Be15}). \citetalias{Be15} use a Fisher matrix analysis to compare the errorbar on $\sig(\om/0.3)^{0.5}$ from the two covariance estimates, and find agreement within the noise expected from the finite number of simulations, with a larger errorbar when using the mock covariance. We believe the analytic halo model approach is a very promising one, which, with further validation (for example investigating the effect of not including the exact survey geometry), has the potential to relax the requirement of producing thousands of mock surveys for future, larger weak lensing datasets. For this study, we believe that the mock covariance, although noisy, is the more reliable and conservative option. 
We apply the correction factor to the inverse covariance described in \citet{hartlap2007}.

The analysis in this paper neglects the cosmology dependence of the covariance, which as outlined in \citet{eifler2009}, can substantially impact parameter constraints, depending on the depth and size of the survey. \citetalias{kilbinger13} find this effect to be small for CFHTLenS and since our data is shallower, we are confident that the cosmology-independent noise terms dominate our statistical error budget.
However, we note that in regions of cosmological parameter space far from the fiducial cosmology assumed for the covariance i.e. in the extremes of the ‘banana’ in e.g. Figure \ref{fig:fid}, the reported uncertainties will be less reliable.

\subsection{Photometric redshift estimates}
\label{sec:photoz}

The photometric redshifts used in this work are described in \citetalias{Bo15}. They compare four  methods: Skynet \citep{Bonnett2015,Graff2013},
TPZ \citep{Carrasco13}, ANNz2 \citep{sadeh2015}
 and BPZ \citep{benitez98}. These methods performed well amongst a more extensive list of methods tested in \citet{sanchez14}. The first three are machine learning methods and are trained on a range of spectroscopic data; the fourth is a template-fitting method, empirically calibrated relative to simulation results from \citet{Chang2014bccufig} and \citet{leistedt15}. The validation details are described in \citetalias{Bo15}, including a suite of tests of the performance of these codes with respect to spectroscopic samples, simulation results, COSMOS photo-zs \citep{ilbert2009}, and relative to each other. They conclude that the photometric redshift estimates of the $n(z)$ of the source galaxies are accurate to within an overall additive shift of the mean redshift of the $n(z)$ with an uncertainty of 0.05. The fiducial photometric redshift method is chosen to be Skynet, as it performed best in tests, but in Section~\ref{subsec:nz_errs} we show the impact of switching to the other methods.

\section{Fiducial cosmological constraints}
\label{sec:fid_constraints}

In this Section we present our headline 
DES SV cosmology results from the fiducial data vector, marginalising over a fiducial set of systematics and cosmology parameters. In the later sections we examine the robustness of our results to various changes of the data vector and modelling of systematics. 
   
We evaluate the likelihood of the data from the two-point estimates and covariances presented in \citetalias{Be15} and the corresponding theoretical predictions, described in Section \ref{sec:choice_2pt} assuming that the 
estimates are drawn from a multi-variate Gaussian distribution.
Key results for this paper have been calculated with
two separate pipelines:
the {\sc CosmoSIS}\footnote{\texttt{https://bitbucket.org/joezuntz/cosmosis}} \citep{cosmosis} 
and {\sc CosmoLike}~\citep{cosmolike} 
frameworks. The constraints from these independent pipelines agree extremely well and thus are not shown separately. {\sc CosmoLike}~uses the \citet{EisensteinHu} prescription for the linear matter power spectrum $\pofk$, and {\sc CosmoSIS} uses {\sc Camb} \citep{Lewis2000}. For a vanilla \lcdm cosmology ($\om=0.3$, $\sig=0.8$, $n_s=0.96$, $h=0.7$), we find theory predictions using {\sc Camb} and \citet{EisensteinHu} differ by at most 1\% for the scales and redshifts we use. For the increased statistical power of future datasets, differences of this order will not be acceptable.

\begin{figure}
\includegraphics[width=\columnwidth]{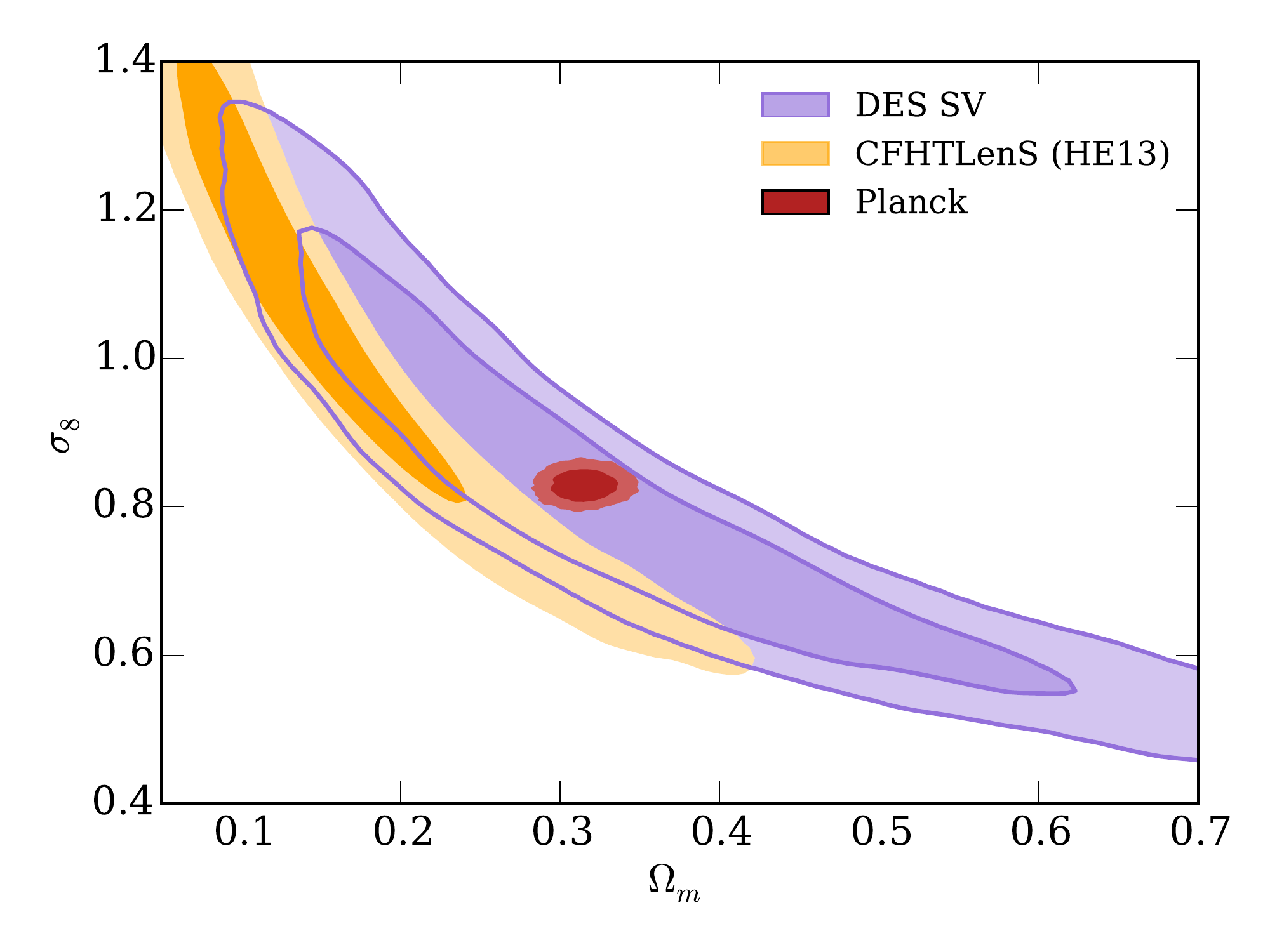}
\caption{
Constraints on the amplitude of fluctuations $\sig$ and the matter density $\om$ from DES SV cosmic shear (purple filled/outlined contours) compared with constraints from Planck (red filled contours) and CFHTLenS (orange filled, using the correlation functions and covariances presented in \citet{heymans13}, and the `original conservative scale cuts' described in Section \ref{subsec:otherlensing}). DES SV and CFHTLenS are marginalised over the same astrophysical systematics parameters and DES SV is additionally marginalised over uncertainties in photometric redshifts and shear calibration. Planck is marginalised over the 6 parameters of \lcdm (the 5 we vary in our fiducial analysis plus $\tau$). 
The DES SV and CFHTLenS constraints are marginalised over wide flat priors on $n_s$, $\Omega_{\rm b}$ and $h$ (see text), 
assuming a flat universe. For each dataset, we show contours which encapsulate 68\% and 95\% of the probability, as is the case for subsequent contour plots.
}
\label{fig:fid}
\end{figure}

The fiducial data vector is the real-space shear--shear angular correlation function $\xi_{\pm}(\theta)$ measured in three redshift bins (hereafter bins 1, 2, 3, with ranges of $0.3<z<0.55$, $0.55<z<0.83$ and $0.83<z<1.3$, and galaxies assigned to bins according the mean of their photometric redshift probability distribution function) including cross-correlations, as shown in Figure \ref{fig:xi_measurements}. The 
data vector
initially includes galaxy pairs with separations between 2 and 300 arcmin (although many of these pairs are excluded by the scale cuts described in Section \ref{subsec:scales}).
We focus mostly on placing constraints on the matter density of the Universe, $\om$, and $\sig$, defined as the rms 
mass density fluctuations in 8 Mpc$/h$ spheres at the present day, as predicted by linear theory.

We marginalise over wide flat priors $0.05<\om<0.9$, $0.2<\sig<1.6$, $0.2<h<1$, $0.01<\Omega_{\rm b}<0.07$ and $0.7 < n_s <1.3$, assuming a flat Universe, and thus we vary 5 cosmological parameters in total. The priors were chosen to be wider than the constraints in a variety of existing Planck chains.
In practice the results are very similar to those with these parameters fixed, due to the weak dependence of cosmic shear on these other parameters. We use a fixed neutrino mass of 0.06 eV.

We summarise our systematics treatments below:\\
{\bf(i) Shear calibration:}
For each redshift bin, we marginalise over a single free parameter to account for shear measurement uncertainties: the predicted data vector is modified to account for a potential unaccounted multiplicative bias as $\xi^{ij} \rightarrow (1+m_i)(1+m_j)\xi^{ij}$. 
We place a separate Gaussian prior on each of the 
three $m_i$ parameters. Each is centred on 0 and of width 0.05, as advocated by \citetalias{J15}. See Section \ref{subsec:shear_cal} for more details.\\ 
{\bf(ii) Photometric redshift calibration:}
Similarly, we marginalise over one free parameter per redshift bin to describe photometric redshift calibration uncertainties. We allow for an independent shift of the estimated photometric redshift distribution $n_i(z)$ in redshift bin $i$ i.e. $n_i(z) \rightarrow n_i(z-\delta z_i)$. 
We use independent Gaussian priors on 
each of the three $\delta z_i$ values of width 0.05 as recommended by 
\citetalias{Bo15}.  See Section \ref{subsec:nz_errs} for more details.\\
{\bf(iii) Intrinsic alignments:}
We assume an unknown amplitude of the intrinsic alignment signal and marginalise over this single parameter, assuming the non-linear alignment model of
\citet{BK07}. See Section \ref{subsec:IAs} for more details of our implementation and tests on the sensitivity of our results to intrinsic alignment model choice. \\
{\bf(iv) Matter power spectrum:}
 We use $\halofit$~ \citep{smith03}, with updates from \citet{takahashi2012} to model the non-linear matter power spectrum, and refer to this prescription simply as `\halofit' henceforth.
The range of scales for the fiducial data vector is chosen to reduce the bias from theoretical uncertainties in the non-linear matter power spectrum to a level which is not significant given our statistical uncertainties (see Sections \ref{subsec:scales} and \ref{subsec:nonlin}, and Table \ref{table:scales} for the minimum angular scale for each bin combination). \\
We thus marginalise over $3+3+1=7$ nuisance parameters characterising potential biases in the shear calibration, photometric redshift estimates and intrinsic alignments respectively.

\begingroup
\squeezetable
\begin{table*}
\begin{tabular}{lcccc}
\hline
Model & $S_8 \equiv \sigma_8 (\Omega_m/0.3)^{0.5}$ & Mean Error & $\alpha$ & $\sigma_8 (\Omega_m/0.3)^\alpha$ \\
\hline
{\it Primary Results} &&&& \vspace{0.15cm}  \\
Fiducial DES SV cosmic shear & $0.812^{+0.059}_{-0.060}$ & 0.059 & $0.478$ & $0.811^{+0.059}_{-0.060}$ \\
No photoz or shear systematics & $0.809^{+0.051}_{-0.040}$ & 0.046 & $0.439$ & $0.806^{+0.051}_{-0.051}$ \\
No systematics & $0.775^{+0.045}_{-0.041}$ & 0.043 & $0.462$ & $0.775^{+0.046}_{-0.041}$ \\
\hline
{\it Data Vector Choice} &&&& \vspace{0.15cm} \\
No tomography & $0.726^{+0.117}_{-0.137}$ & 0.127 & $0.513$ & $0.730^{+0.117}_{-0.138}$ \\
No tomography or systematics & $0.719^{+0.063}_{-0.053}$ & 0.058 & $0.487$ & $0.716^{+0.060}_{-0.060}$ \\
$\xi$-to-$C_{\ell}$ bandpowers, no tomo. or systematics & $0.744^{+0.075}_{-0.055}$ & 0.065 & $0.459$ & $0.739^{+0.089}_{-0.055}$ \\
PolSpice-$C_{\ell}$ bandpowers, no tomo. or systematics & $0.729^{+0.094}_{-0.058}$ & 0.076 & $0.518$ & $0.732^{+0.084}_{-0.061}$ \\
\hline
{\it Shape Measurement} &&&& \vspace{0.15cm} \\
Without shear bias marginalisation & $0.812^{+0.054}_{-0.054}$ & 0.054 & $0.492$ & $0.811^{+0.054}_{-0.054}$ \\
{\sc Im3shape} shears & $0.875^{+0.088}_{-0.075}$ & 0.082 & $0.579$ & $0.862^{+0.089}_{-0.075}$ \\
\hline
{\it Photometric Redshifts} &&&& \vspace{0.15cm} \\
Without photo-z bias marginalisation & $0.809^{+0.055}_{-0.054}$ & 0.054 & $0.486$ & $0.808^{+0.054}_{-0.054}$ \\
TPZ photo-zs & $0.814^{+0.059}_{-0.059}$ & 0.059 & $0.499$ & $0.814^{+0.059}_{-0.059}$ \\
ANNZ2 photo-zs & $0.827^{+0.060}_{-0.060}$ & 0.060 & $0.483$ & $0.826^{+0.060}_{-0.059}$ \\
BPZ photo-zs & $0.848^{+0.063}_{-0.064}$ & 0.063 & $0.474$ & $0.845^{+0.063}_{-0.064}$ \\
\hline
{\it Intrinsic Alignment Modelling} &&&& \vspace{0.15cm} \\
No IA modelling & $0.770^{+0.053}_{-0.053}$ & 0.053 & $0.477$ & $0.769^{+0.054}_{-0.053}$ \\
Linear alignment model & $0.799^{+0.063}_{-0.054}$ & 0.059 & $0.479$ & $0.799^{+0.062}_{-0.053}$ \\
Tidal alignment model & $0.810^{+0.061}_{-0.060}$ & 0.060 & $0.494$ & $0.810^{+0.060}_{-0.060}$ \\
Marginalised over redshift power law & $0.720^{+0.153}_{-0.153}$ & 0.153 & $0.449$ & $0.723^{+0.145}_{-0.146}$ \\
Marginalised over redshift power law with $A>0$ & $0.808^{+0.058}_{-0.058}$ & 0.058 & $0.493$ & $0.807^{+0.058}_{-0.057}$ \\
\hline
{\it High-k power spectrum} &&&& \vspace{0.15cm} \\
Without small-scale cuts & $0.819^{+0.068}_{-0.062}$ & 0.065 & $0.487$ & $0.819^{+0.066}_{-0.061}$ \\
OWLS AGN $P(k)$ & $0.820^{+0.060}_{-0.061}$ & 0.061 & $0.485$ & $0.819^{+0.060}_{-0.061}$ \\
OWLS AGN $P(k)$ w/o small-scale cuts & $0.838^{+0.069}_{-0.059}$ & 0.064 & $0.484$ & $0.838^{+0.067}_{-0.058}$ \\
\hline
{\it Other lensing data} &&&& \vspace{0.15cm} \\
{\bf CFHTLenS (H13) original conservative scales} & $0.710^{+0.040}_{-0.034}$ & 0.037 & $0.497$ & $0.712^{+0.040}_{-0.034}$ \\
CFHTLenS (H13) modified conservative scales & $0.692^{+0.044}_{-0.033}$ & 0.038 & $0.474$ & $0.704^{+0.041}_{-0.031}$ \\
CFHTLenS (H13) + DES SV & $0.744^{+0.035}_{-0.031}$ & 0.033 & $0.487$ & $0.747^{+0.034}_{-0.028}$ \\
CFHTLenS (K13) all scales & $0.738^{+0.055}_{-0.032}$ & 0.043 & $0.480$ & $0.739^{+0.066}_{-0.031}$ \\
CFHTLenS (K13) original conservative scales & $0.596^{+0.080}_{-0.073}$ & 0.077 & $0.602$ & $0.622^{+0.077}_{-0.071}$ \\
CFHTLenS (K13) modified conservative scales & $0.671^{+0.067}_{-0.061}$ & 0.064 & $0.562$ & $0.688^{+0.055}_{-0.047}$ \\
Planck Lensing & $0.820^{+0.100}_{-0.141}$ & 0.121 & $0.241$ & $0.799^{+0.027}_{-0.030}$ \\
\hline
{\it Planck 2015 Combination/Comparison} &&&& \vspace{0.15cm} \\
{\bf Planck (TT+LowP)} & $0.850^{+0.024}_{-0.024}$ & 0.024 & $-0.021$ & $0.829^{+0.014}_{-0.015}$ \\
Planck (TT+LowP)+DES SV & $0.848^{+0.022}_{-0.021}$ & 0.022 & $-0.002$ & $0.829^{+0.013}_{-0.014}$ \\
Planck (TT+EE+TE+Low TT) & $0.861^{+0.020}_{-0.020}$ & 0.020 & $0.321$ & $0.856^{+0.018}_{-0.019}$ \\
Planck (TT+LowP+Lensing) & $0.825^{+0.017}_{-0.017}$ & 0.017 & $0.098$ & $0.817^{+0.009}_{-0.009}$ \\
Planck (TT+LowP+Lensing)+ext & $0.824^{+0.013}_{-0.013}$ & 0.013 & $0.098$ & $0.817^{+0.010}_{-0.009}$ \\
\hline
\end{tabular}

\caption{
68\% confidence limits
on $S_8 \equiv \sig(\om/0.3)^{0.5}$ in $\Lambda$CDM for various assumptions 
in the DES SV analysis, compared to CFHTLenS and Planck and combined with various datasets. In the first column the power law index from the fiducial case, $0.478$, is rounded to $0.5$ and used for all variants. The second column shows the symmetrised error bar on $S_8$ for ease of comparison between rows. 
In the third column we show the fitted power law index $\alpha$ for each variant, and in the final column we show the constraint on $\sig(\om/0.3)^{\alpha}$, where the value of $\alpha$ is fixed to the value given in the third column, separately for each variant.  A graphical form of the first column is shown in Figure \ref{fig:s8om_values}.
}
\label{table:om_s8}
\end{table*}
\endgroup

\begin{figure*}
\centerline{\includegraphics[width=14cm]{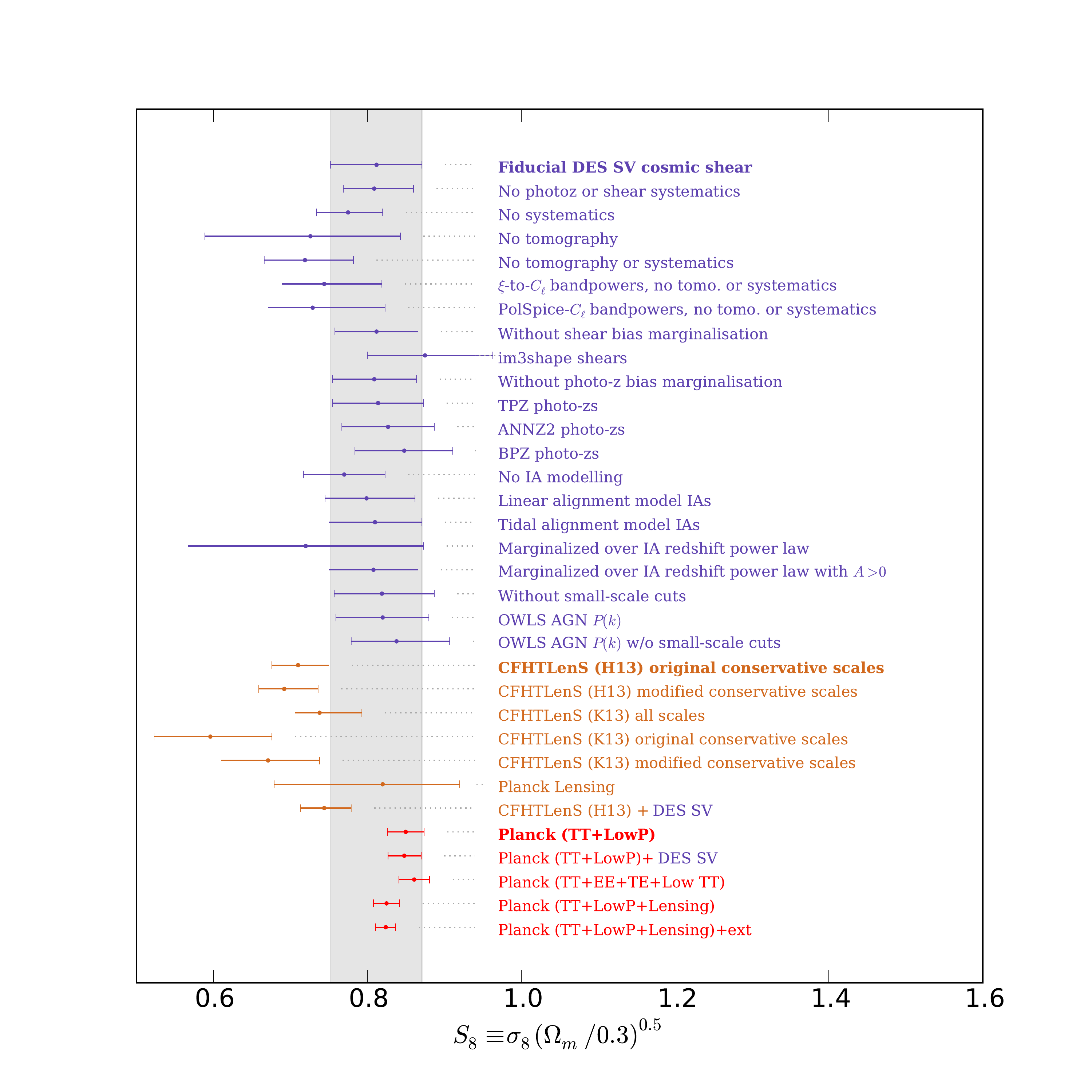}}
\caption{
Graphical illustration of the 68\% confidence limits on $S_8\equiv\sig(\om/0.3)^{0.5}$ values given in Table \ref{table:om_s8}, showing the robustness of our results (purple) and comparing with the CFHTLenS and Planck lensing results (orange) and Planck (red). The grey vertical band aligns with the fiducial constraints at the top of the plot.
Note that Planck lensing in particular, and other non-DES lensing measurements optimally constrain a different quantity than shown above e.g. see the second and third columns of Table \ref{table:om_s8}.
}
\label{fig:s8om_values}
\end{figure*}

Figure \ref{fig:fid} shows our main DES SV cosmological constraints in the $\om-\sig$ plane, from the fiducial data vector and systematics treatment, compared to those from CFHTLenS and Planck. For the CFHTLenS constraints, we use the same six redshift bin data vector and covariance as \citetalias{heymans13}, but apply the conservative cuts to small scales used as a consistency test in that work (for $\xip$ we exclude angles $<3'$ for redshift bin combinations involving the lowest two redshift bins, and for $\xim$, we exclude angles $<30'$ for bin combinations involving the lowest four redshift bins, and angles $<16'$ for bin combinations involving the highest two redshift bins). We see that in this plane, our results are midway between the two datasets and are compatible with both. We discuss this further in Section~\ref{sec:comparisons}. 

Using the MCMC chains generated for Figure \ref{fig:fid} 
we find the best fit power law $\sig(\om/0.3)^{\alpha}$ to describe the degeneracy direction in the $\sig$, $\om$ plane (we estimate $\alpha$ using the covariance of the samples in the chain in $\rm{log}\sig-\rm{log}\om$ space). We find $\alpha=0.478$ and so use a fiducial value for $\alpha$ of 0.5 for the remainder of the paper \footnote{We would advise caution when using $S_8$ to characterise the DES SV constraints instead of a full likelihood analysis - $S_8$ is sensitive to the tails of the probability distribution, and also weakly depends on the priors used on the other cosmological parameters}.
We find a constraint perpendicular to the degeneracy direction of
\begin{equation}
S_8 \equiv \sig(\om/0.3)^{0.5}= 0.81 \pm 0.06 \quad (68\%).
\end{equation}
Because of the strong degeneracy, the marginalised 1d constraints on either $\om$ or $\sig$ alone are weaker; we find $\om=0.36^{+0.09}_{-0.21}$ and $\sig=0.81^{+0.16}_{-0.26}$.
In Table \ref{table:om_s8} we also show other results which are discussed in the later sections, including variations of the DES SV analysis (see Sections ~\ref{sec:choice_2pt} and ~\ref{sec:constraints}) and combinations with CFHTLenS and Planck (see Section~\ref{sec:comparisons}).

\begin{figure}
\includegraphics[width=\columnwidth]{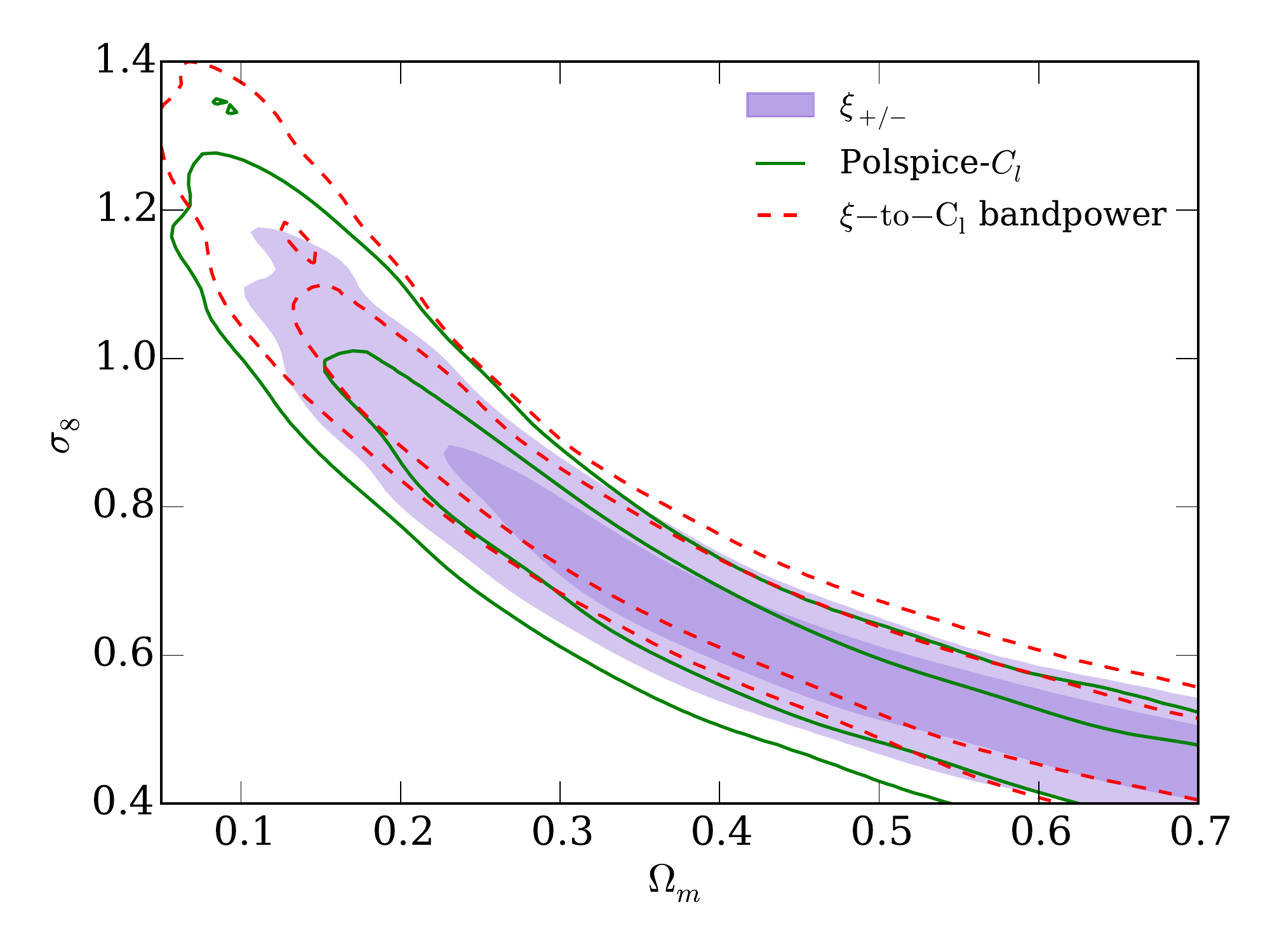}
\caption{
Comparison of constraints on $\sig$ and $\om$ for various choices of data vector: $\xi_{\pm}$ with no tomography or systematics (purple filled), $C^{ij}_{\ell}$ bandpowers (dashed red lines) and PolSpice-$C_{\ell}$ bandpowers (solid green lines) (both with no tomography or systematics). We do not show our fiducial constraints, or Planck, since we have not marginalised over systematics for the constraints shown here, so agreement is not necessary or meaningful (although Table \ref{table:om_s8} suggests there is reasonable agreement). 
}
\label{fig:2pt-estimators}
\end{figure}

For comparison with other constraints we also investigated the impact of ignoring shear measurement and photometric redshift uncertainties and find that the central value of $S_8$ changes negligibly, and the error bar decreases by $\sim$20\% (see Table~\ref{table:om_s8} for details). 

In Table~\ref{table:om_s8} we also show results ignoring all systematics. This is the same as the ``No photoz or shear systematics'' case but additionally ignoring intrinsic alignments, so that only the other cosmological parameters are varied. The central value shifts down by 0.037 and the error bar is reduced by 27\% compared to the fiducial case. Therefore the systematics contribute almost half (in quadrature) of our total error budget, and further effort will be needed to reduce systematic uncertainties if we are to realise a significant improvement in the constraints (from shear--shear correlations alone) with larger upcoming DES samples.

\section{Choice of data vector and scales used}
\label{sec:data_vector}

In this Section we consider the impact of the choice of two-point statistic on the cosmological constraints, and investigate how our fiducial estimators are affected by the choice of 
angular scales used.

\subsection{Choice of two-point statistic}
\label{sec:choice_2pt}

\citetalias{Be15} present results for a selection of two-point statistics -- see that work, and references therein for more detailed description of the statistics and their estimators.  For an overview of the theory presented here see \citet{BartelmannSchneider2001}.

The statistics can all be described as weighted integrals over 
the weak lensing convergence power spectrum at angular wavenumber $\ell$, 
$C^{ij}_{\ell}$, of tomographic redshift bin $i$ and $j$, which can be related to the matter power spectrum, $\pofk$, by the Limber approximation
\be
C ^{ij} _\ell = \frac{9H_0^4 \om^2}{4c^4} \int_0^{\chi_\mr h} 
\mr d \chi \, \frac{g^{i}(\chi) g^{j}(\chi)}{a^2(\chi)} \pd \left(\frac{\ell}{f_K(\chi)},\chi \right) \,,
\ee
where $\chi$ is the comoving radial distance, $\chi_\mr h$  is the comoving distance of the horizon, $a(\chi)$ is the scale factor, and $f_K(\chi)$ the comoving angular diameter distance. We assume a flat universe ($f_K(\chi)=\chi$) hereafter. The lensing efficiency $g^{i}$ is defined as an integral over the redshift distribution of source galaxies $n(\chi(z))$ in the $i^\mr{th}$ redshift bin:
\be
g^{i}(\chi) = \int_\chi^{\chi_{\mr h}} \mr d \chi' n_{i} (\chi') \frac{f_K (\chi'-\chi)}{f_K (\chi')} \,,
\ee
Our fiducial statistics, the real space correlation functions, $\xipm(\theta)$, are weighted integrals of the angular power spectra:
\be
\label{eq:2PCFshear}
\xipm^{ij} (\vt) = \frac{1}{2 \pi} \int \d \ell \,  \ell \, J_{0/4} (\ell \vt) \, C ^{ij}_{\ell}  \,, \\
\ee
where $J_{0/4}$ is the Bessel function of either $0^\mr{th}$ or $4^\mr{th}$ order. $\xipm$ have the advantage of being straightforward to estimate from the data, 
whereas the $C^{ij}_{\ell}$s require more processing but are a step closer to the theoretical predictions.
An advantage of using $C^{ij}_{\ell}$ is that the signal is split into two parts, E- and B-modes, the latter of which is expected to be very small for cosmic shear. The cosmic shear signal is concentrated in the E-mode because to first order the shear signal is the gradient of a scalar field. The B-mode can therefore be used as a test of systematics as
discussed in \citetalias{J15} and \citetalias{Be15}.

\begin{figure}
\includegraphics[width=\columnwidth]{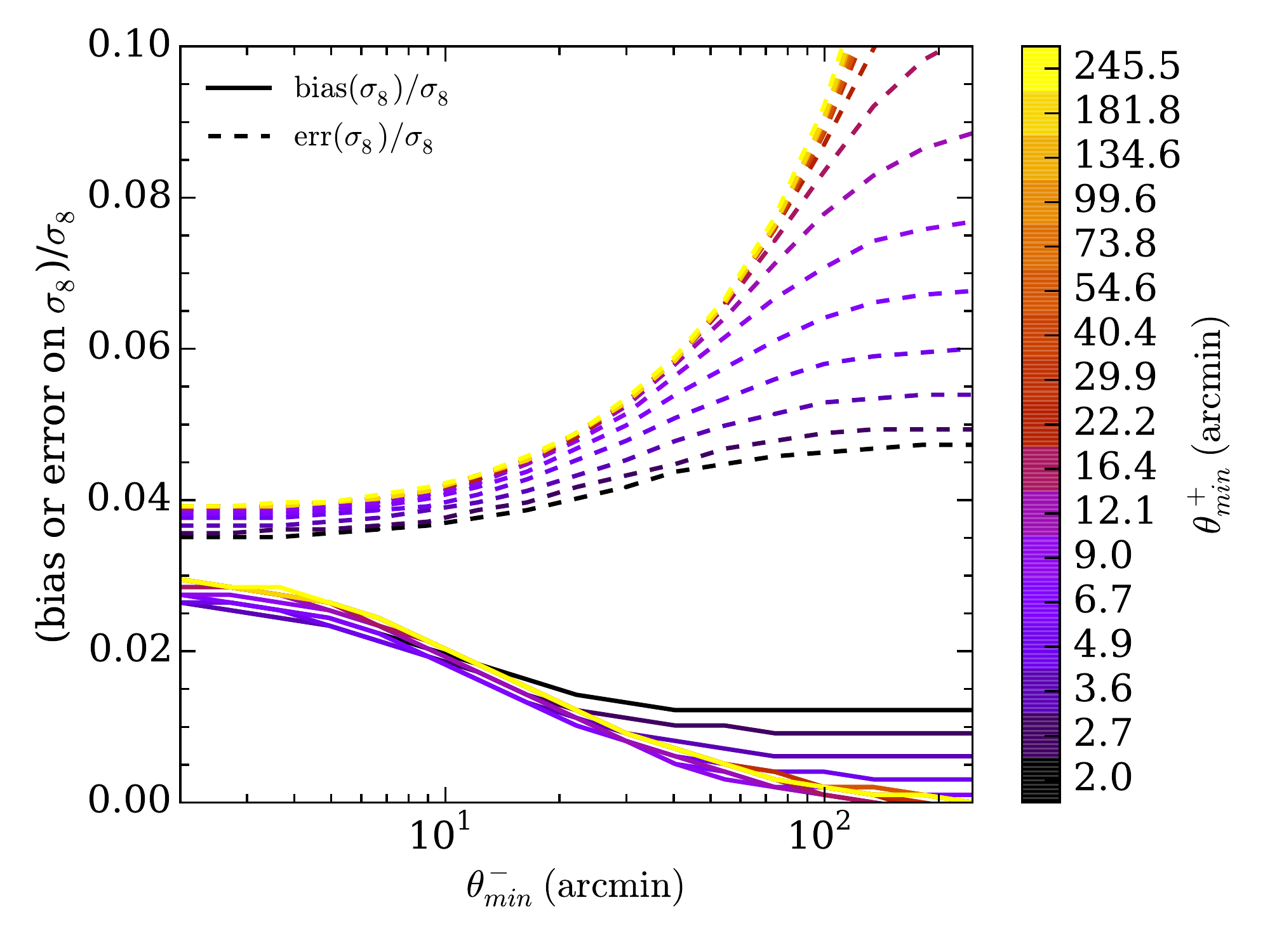}
\caption{The fractional bias on $\sig$ due to ignoring an OWLS AGN baryon model (solid lines) compared to the statistical uncertainty on $\sig$ (dashed lines) as a function of minimum scale used for $\xim$ ($\theta_{\rm{min}}^-$, x-axis) or $\xip$ ($\theta_{\rm{min}}^+$, colours).
Whereas the statistical error is minimised by using small scales, the bias is significant for $\theta_{\rm{min}}^-<30'$ and $\theta_{\rm{min}}^+<3'$.}
\label{fig:scales}
\end{figure}

\citetalias{Be15} also implement the method of \citet{becker2014} which uses linear combinations of $\xipm(\theta)$ to estimate fourier space bandpowers of $C^{ij}_{\ell}$.
Also presented are PolSpice \citep{szapudi2001} estimates of the $C^{ij}_{\ell}$s 
from pixelised shear 
maps using the pseudo-$C_{\ell}$ estimation process, which corrects the spherical harmonic transform values for the effect of the survey mask (see \citet{hikage2011} for the first implementation for cosmic shear). 
For simplicity we do not perform a tomographic analysis using these estimators. To compare cosmological constraints with these different estimators we do not marginalise over any systematics, to enable a more conservative comparison between them. (Note that marginalising over intrinsic alignments inflates the errors of non-tomographic analyses as described in Section~\ref{subsec:IAs}).  
Figure \ref{fig:2pt-estimators} 
shows constraints from the different estimators, and we see that the three are in good agreement. 
A more detailed comparison can be made using the numbers in Table \ref{table:om_s8}, which are shown graphically in Figure \ref{fig:s8om_values}.
The relevant lines for comparison are the ``No tomography or systematics'' line which uses the fiducial $\xipm$ data vector, and the two $C_{\ell}$ bandpower lines. The uncertainties are similar between these methods, and the PolSpice-$C_\ell$ constraints are shifted to slightly lower $S_8$, though are consistent with constraints based on the $\xipm$ approach. Although we find the qualitative agreement between the constraints from the different estimators encouraging, we note that testing on survey simulations would be required to make a quantitative statement about the level of agreement.

\subsection{Choice of scales}\label{subsec:scales}

All the two-point statistics discussed thus far involve a mixing of physical scales:
it is clear from Eq. \ref{eq:2PCFshear} that $\xipm$ at a given real space angular scale uses information from a range of 
angular wavenumbers $\ell$, while $C_{\ell}$ itself uses information from a range of physical scales $k$ in the matter power spectrum $\pofk$. 
In Section \ref{subsec:nonlin} we discuss some of the difficulties in producing an accurate theoretical estimate of $\pofk$ for high $k$ (small physical scales).
In this work, we aim to null the effects of this theoretical uncertainty by cutting small angular scales from our data vector, since using scales where the theoretical prediction is inaccurate can bias the derived cosmological constraints, 
mostly due to unknown baryonic effects on clustering.

Figure \ref{fig:scales} demonstrates the impact of errors in the matter power spectrum prediction on estimates of $\sigma_8$ from a non-tomographic analysis. 
In this figure we estimate the potential bias on $\sig$ as that which would arise from ignoring the presence of baryonic effects; as a specific model for these effects we use the OWLS AGN simulation \citep{schaye10}. See Section \ref{subsec:nonlin} for more details, in particular Eq. \ref{eq:owls} for the implementation of the AGN model. 
For a given angular scale $\xim$ is more affected than $\xip$: for example the fractional bias when using all scales in $\xim$, but none in $\xip$ $(\theta_{\rm{min}}^-=2', \theta_{\rm{min}}^+=245.5')$ is $\approx0.03$ whereas the bias when using all scales in $\xip$, but none in $\xim$ ($\theta_{\rm{min}}^+=2', \theta_{\rm{min}}^-=245.5)$ is $\approx0.015$.
For the non-tomographic case, we 
use a minimum angular scale 
of 
3 arcminutes for $\xip$, and 30 arcminutes for $\xim$, 
because on these angular scales the bias is $<25\%$ of the statistical uncertainty on $\sig$ (with no other parameters marginalised). 


For the tomographic case, we now need to choose a minimum scale for xi+ and xi- for each of the redshift bin combinations - i.e. 12 parameters. Hence a procedure analagous to that based on Figure 5 is non-trivial.    We instead use a more general (but probably non-optimal) prescription in which we cut angular bins that change significantly when we change the model for the non-linear matter power spectrum. We remove data points where the theoretical prediction changes by more than 5\% when the nonlinear matter power spectrum is switched from the fiducial to either that predicted from the FrankenEmu\footnote{http://www.hep.anl.gov/cosmology/CosmicEmu/emu.html} code (based on the Coyote Universe Simulations described in \citet{heitmann2014}, and extended at high $k$ using the `CEp' presciption from \citet{harnois14}), or to the OWLS AGN model (the baryonic model used in Figure 5). We believe 5\% is a reasonable (but again, probably not optimal) choice, since on these nonlinear scales, the signal is proportional to $\sig^3$, so a 5\% prediction error would result in a $\sig$ error of order 0.05/3 i.e. below our statistical uncertainties. The inferred biases for the non-tomographic $\xipm$ shown in Figure~\ref{fig:scales} suggest similar angular cuts. The results of these cuts are summarised in Table~ \ref{table:scales}. We demonstrate the effectiveness of these cuts in producing robust robust constraints, and discuss other methods of dealing with non-linear scales in Section \ref{subsec:nonlin}.

We limit the large scales in $\xip$ to $<60$ arcmin, since the large scales in $\xip$ are highly correlated, and we have verified that little is gained in signal-to-noise by including larger scales. Furthermore, including these larger scales would also increase the number of data points, increasing the noise in the covariance matrix, and degrading our parameter constraints.

\begin{table}
\centering
\begin{tabular}{lcc}
\hline\hline
Redshift bin combination & $\theta_{min}(\xip)$ &  $\theta_{min}(\xim)$ \\
\hline
(1,1) & 4.6 & 56.5 \\
(1,2) & 4.6 & 56.5 \\
(1,3) & 4.6 & 24.5 \\
(2,2) & 4.6 & 24.5 \\
(2,3) & 2.0 & 24.5 \\
(3,3) & 2.0 & 24.5 \\
\hline\hline
\end{tabular}
\caption{Scale cuts for tomographic
shear two point functions 
$\xipm$ 
using the prescription described in the text. 
}
\label{table:scales}
\end{table}

\section{Robustness to systematics}
\label{sec:constraints}

We now examine the robustness of our fiducial constraints to assumptions made about the main systematic uncertainties for cosmic shear. In each subsection we consider the impact of ignoring the systematic in question, and examine alternative prescriptions for the input data or modelling.

\begin{figure}
\includegraphics[width=\columnwidth]{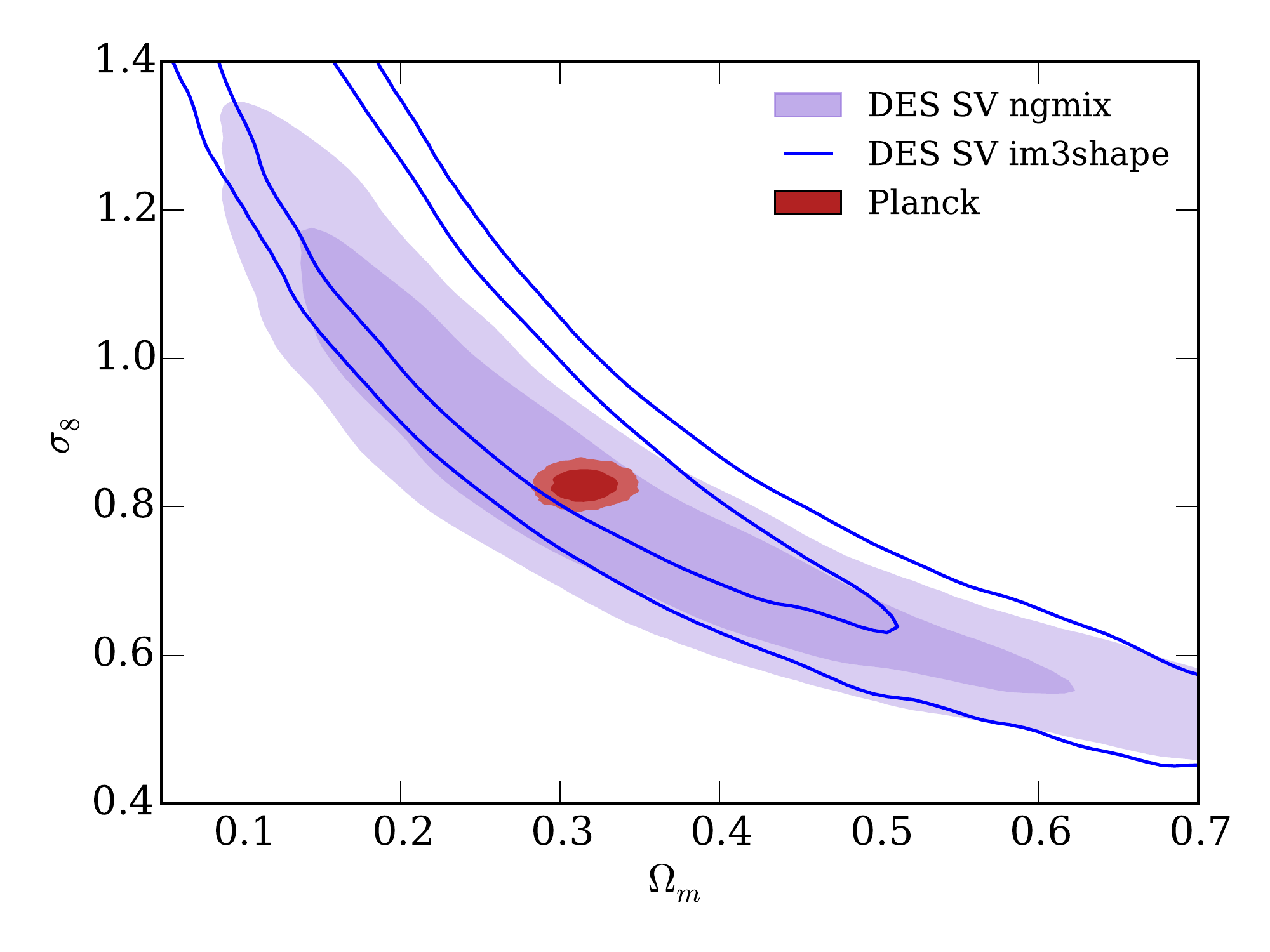}
\caption{
Robustness to assumptions about shear measurement. Shaded purple (fiducial case): \ngmix, with one shear mulitiplicative bias parameter m for each of the 3 tomographic redshift bins, with an independent Gaussian prior on each $m_i$ with $\sigma=0.05$. Solid blue lines: \imshape~with the same assumptions. 
Planck is shown in red.
}
\label{fig:shears}
\end{figure}

\subsection{Shear calibration}
\label{subsec:shear_cal}

The measurement of galaxy shapes at the accuracy required for cosmic shear is a notoriously hard problem. The raw shapes in our two catalogues are explicitly corrected for known sources of systematic bias. This involves either calibration using image simulations in the case of \imshape~or sensitivity corrections in the case of \ngmix~(see \citetalias{J15}). We rely on a number of assumptions and cannot be completely certain the final catalogues carry no residual bias. It is therefore important that  
our model includes the possibility of error in our shape measurements.  As in \citet{jee2013} we marginalise 
over shear measurement uncertainties 
in parameter estimation.

\citetalias{J15} estimate the systematic uncertainty on the shear calibration by comparing the two shape measurement codes to image simulations, and to each other. Following that discussion we include in our model 
a multiplicative uncertainty which is independent in each of the three redshift bins. We thus introduce three free parameters $m_i$ ($i=1,2,3$). 
The predicted data are transformed as
\be
\xi_{\pm \rm{pred}}^{i,j} = (1+m_i)(1+m_j)\xi_{\pm {\rm true}}^{i,j} \;\; 
\ee
for redshift bins $i,j$.

As discussed in \citetalias{J15}, we use a Gaussian prior on the $m_i$ parameters of width 0.05, compared to a 0.06 uniform prior used by \citet{jee2013}. We note that since the $m_i$ are independent, the effective prior on the mean multiplicative bias for the whole sample is less than 0.05.
No systematic shear calibration uncertainties were propagated by CFHTLenS in \citetalias{heymans13} or earlier work (although \citetalias{kilbinger13} did investigate the statistical uncertainty on the shear calibration arising from having a limited calibration sample).
If we neglect this uncertainty and assume that our shape measurement has no errors (fixing $m_i=0$) then our uncertainty on $S_8$ is reduced by 9\% and the central value is unchanged (see the ``Without shear bias marginalisation'' row in Table \ref{table:om_s8} and Figure \ref{fig:s8om_values} for more details).

Figure \ref{fig:shears} shows the result of interchanging the two shear measurement codes, swapping \ngmix~(fiducial) to \imshape. The \imshape~constraints are weaker, because the shapes are measured from a single imaging band (r-band) instead of simultaneously fitting to three bands (r, i, z) as in \ngmix, and \imshape~retains fewer galaxies after quality cuts (in particular the \imshape~catalogue contains around half as many galaxies as \ngmix~in our highest redshift bin). The preferred value of $S_8$ is shifted about $1\sigma$ higher for \imshape~than \ngmix~and the error bar is increased by 38\% (see the ``\imshape~shears'' row in Table \ref{table:om_s8} and Figure \ref{fig:s8om_values}). While we do not expect the constraints from the two shear codes to be identical, since they come from different data selections, the two codes do share many of the same galaxies, and of course probe a common volume. We can estimate the significance of the shift using the mock DES SV simulations detailed in \citetalias{Be15}. Carefully taking into account the overlapping galaxy samples, correlated shape noise and photon noise, and of course the common area, we create an \ngmix~and an \imshape~realisation of our signal for each mock survey. We then compute the difference in the best-fit $\sig$s (keeping all other parameters fixed to fiducial values for computational reasons) for the two signals, and compute the standard deviation of this difference over the 126 mock realisations. We find this difference has a standard deviation of 0.028, compared with the difference in this statistic (the best-fit $\sig$ with all other parameters fixed) on the data of 0.046. We conclude that although this shift is not particularly significant, it could be an indication of shape measurement biases in either catalogue. The decreased statistical errors of future DES analyses will provide more stringent tests on shear code consistency.

\begin{figure}
\includegraphics[width=\columnwidth]{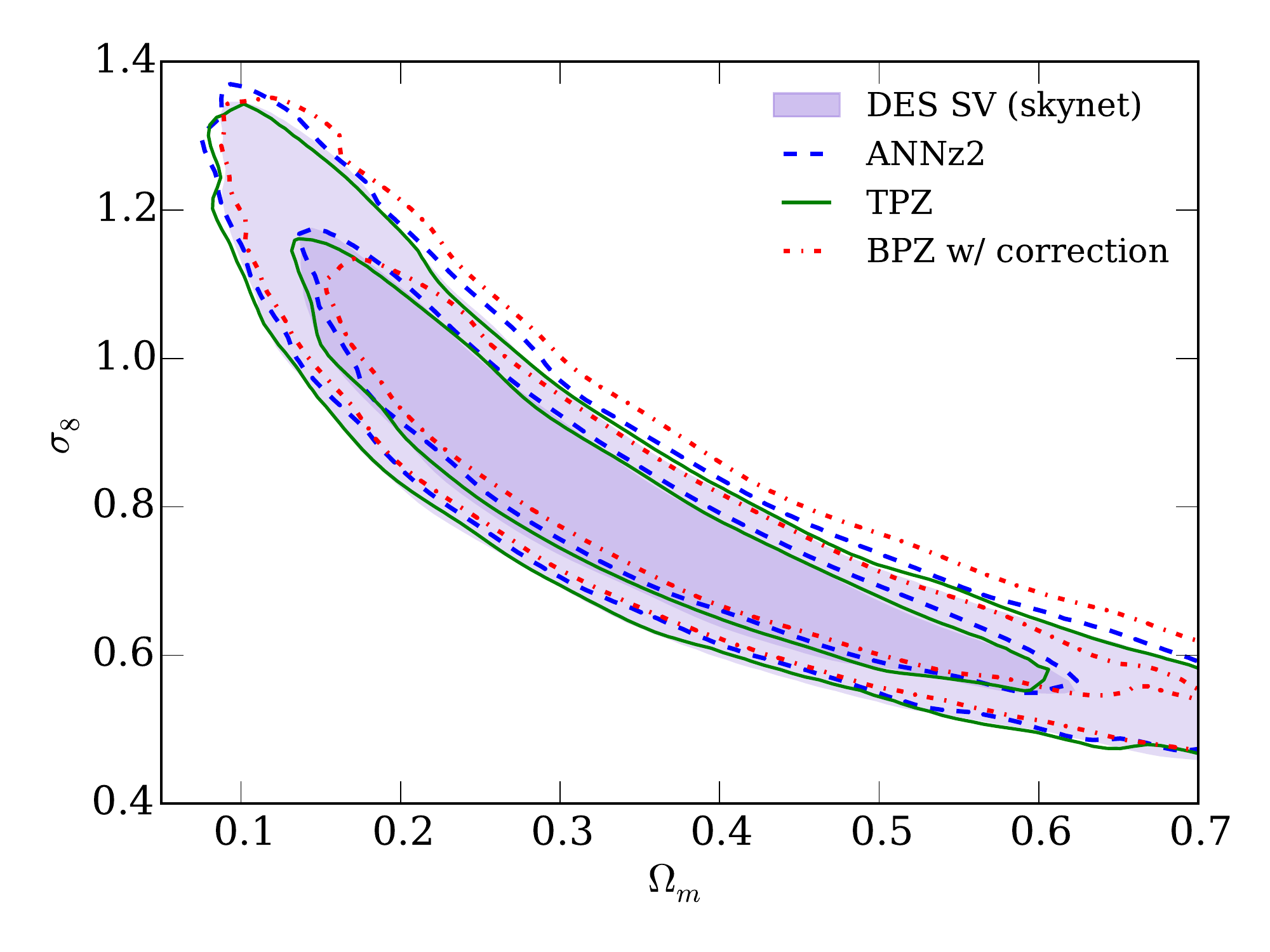}
\caption{
Results using different photoz codes. Purple filled contours: fiducial case (SkyNet). Blue dashed lines: ANNz2. Green solid lines: TPZ. Red dash-dotted lines: BPZ w/ correction. }
\label{fig:photozs}
\end{figure}

\begin{figure*}
\includegraphics[width=\columnwidth]{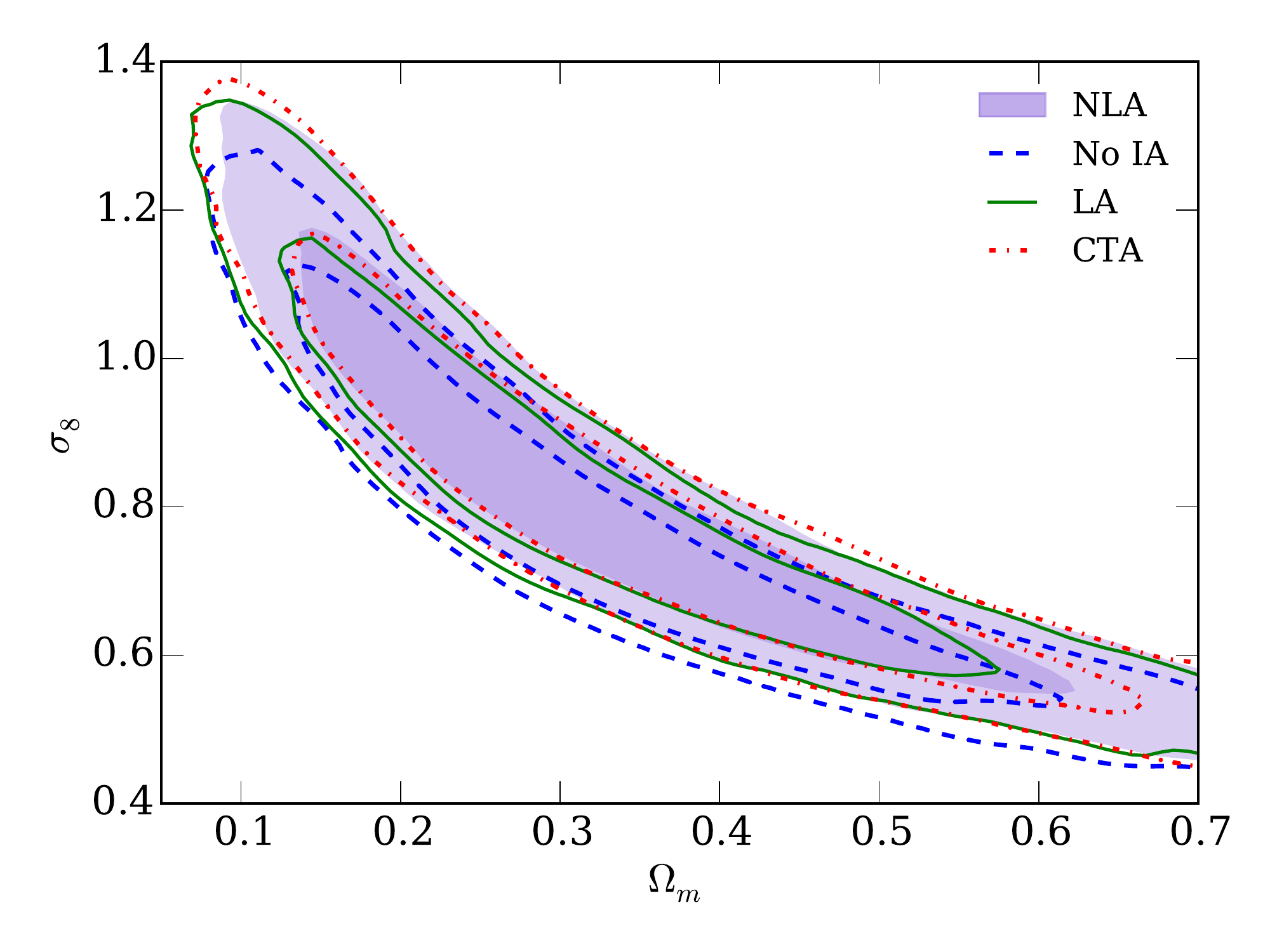}
\includegraphics[width=\columnwidth]{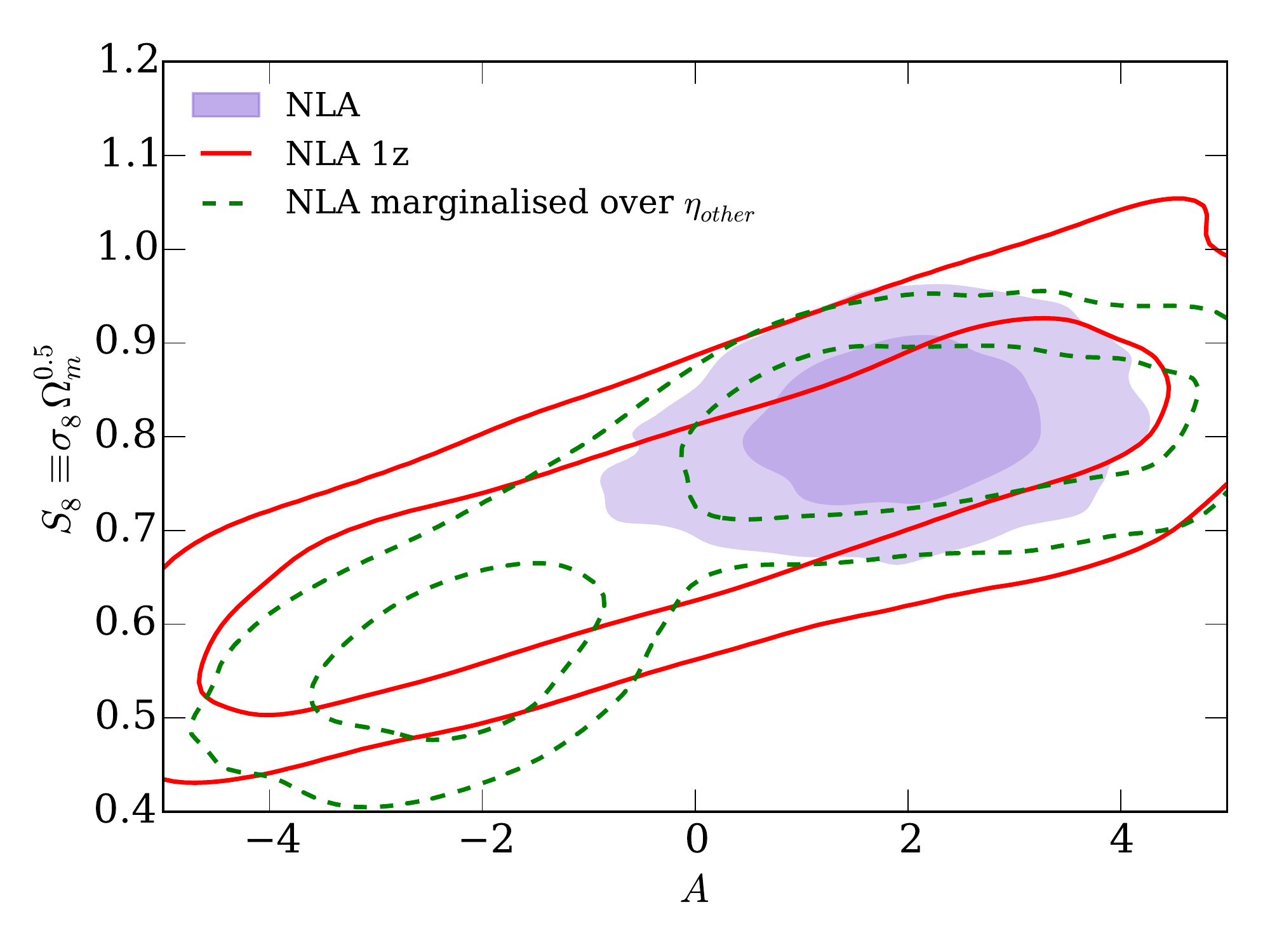}
\caption{Left: Constraints on the clustering amplitude $\sig$ and the matter density $\om$ from DES SV alone. The purple shaded contour shows the constraints when our fiducial NLA model of intrinsic alignments is assumed, the green filled lines shows constraints when the LA model is used, the dot-dashed red lines the CTA model and the blue dashed lines shows constraints when IAs are ignored.
Right: Constraints on $\sig\om^{0.5}$ and the intrinsic alignment amplitude $A$ from DES alone. The purple shaded contour shows the constraints when our fiducial NLA model of intrinsic alignments is assumed with three tomographic bins, the red lines shows constraints, again using our fiducial NLA model, but using only a single redshift bin and the green dashed contour shows our fiducial NLA model, with three tomographic bins, but marginalised over an additional power law in redshift, where the power law index is a free parameter. Note that the treatment of IAs in both panels assumes a prior range for the amplitude $A = [-5,5]$.
}\label{fig:IA_fid_LA_JAB_noIA}
\end{figure*}

\subsection{Photometric redshift biases}
\label{subsec:nz_errs}

In this subsection we investigate the robustness of our constraints to errors in the photometric redshifts. 
As motivated by \citetalias{Bo15}, for our fiducial model we marginalise with a Gaussian prior of width 0.05 over 
three independent photometric redshift calibration bias parameters $\delta z_i$ ($i = 1, 2, 3$) where
\be
n_{i}^{\mathrm{pred}}(z) = n_{i}^{\mathrm{meas}} (z-\delta z_i)
\ee
for redshift bin $i$, where $n_{i}^{\mathrm{meas}}(z)$ is the measured photometric redshift probability distribution and $n_{i}^{\mathrm{pred}}(z)$ is the redshift distribution used in predicting the shear two-point functions (i.e. our model for the true $n_i(z)$ assuming the given $\delta z_i$). This model is discussed further in \citetalias{Bo15} where it is shown to be a reasonably good model for the uncertainties at the current level of accuracy required.

If we neglect photometric redshift calibration uncertainties then the error on $S_8$ is reduced by $\sim$10\% and its value shifts down by $\sim$10\% of the fiducial error bar (see the row labelled ``Without photo-z bias marginalisation'' in Table \ref{table:om_s8} and Figure \ref{fig:s8om_values}). 

In Figure \ref{fig:photozs}
we show the impact of switching between the four photometric redshift estimation codes described in \citetalias{Bo15}. We see excellent agreement between the codes, although as detailed in \citetalias{Bo15}, the machine learning codes are not independent - Skynet, ANNZ2, TPZ are trained on the same spectroscopic data, while an empirical calibration is performed on the template fitting method BPZ using simulation results.
As quantified in Table \ref{table:om_s8} and illustrated in Figure \ref{fig:s8om_values}, the constraint on $S_8$ moved by less than two thirds of the error bar when switching between photometric redshift codes, with the biggest departure occurring for BPZ, which moves to higher $S_8$. A more detailed analysis and validation of the photo-zs using relevant weak lensing estimators and metrics is performed in \citetalias{Bo15} for galaxies in the shear catalogues.

\subsection{Intrinsic alignments}
\label{subsec:IAs}

In this subsection we investigate the effect of assumptions made about galaxy intrinsic alignments (IAs),
by repeating the cosmological analysis with (i) no intrinsic alignments, (ii) a simpler, linear, intrinsic alignment model, (iii)  a more complete tidal alignment model, and (iv) adding a free power law redshift evolution. We also show constraints on the amplitude of intrinsic alignments and show the benefit of using tomography. 
We use the same data vector and likelihood calculation for all models.

It was realised early in the study of weak gravitational lensing \citep{HRH2000,croft00,CKB01,CNP+01} that the unlensed shapes of physically close galaxies may align during galaxy formation due to the influence of the same large-scale gravitational field. This type of correlation was dubbed ``Intrinsic-Intrinsic'', or II. \citet{HS04} then demonstrated that a similar effect can give rise to long-range IA correlations as background galaxies are lensed by the same structures that correlate with the intrinsic shapes of foreground galaxies. This gives rise to a ``Gravitational-Intrinsic'', or GI, correlation. The total measured cosmic shear signal is the sum of the pure lensing contribution and the IA terms:
\begin{equation}
C^{ij}_{\rm obs}(\ell) = C^{ij}_{\rm GG}(\ell) + C^{ij}_{\rm GI}(\ell) + C^{ij}_{\rm IG}(\ell) + C^{ij}_{\rm II}(\ell). 
\end{equation}
Neglecting this effect can lead to significantly biased cosmological constraints \citep{HRH2000,BK07,JMA+11,KRH+12,KEB15}.  

We treat IAs in the ``tidal alignment'' paradigm, which assumes that intrinsic galaxy shapes are linearly related to the tidal field \citep{CKB01}, and thus that the additional $C^{ij}(\ell)$ terms above are integrals over the 3D matter power spectra. It has been shown to accurately describe red/elliptical galaxy alignments \citep{JMA+11,BMS11}. More details of all the IA models considered in this paper can be found in Appendix \ref{sec:IA_appendix}. Within the tidal alignment paradigm, the leading-order correlations define the linear alignment (LA) model. As our fiducial model, we use the ``non-linear linear alignment'' (NLA) model, an ansatz introduced by \citet{BK07}, in which the non-linear matter power spectrum, $P^{\rm nl}_{\delta\delta}(k,z)$, is used in place of the linear matter power spectrum, $P^{\rm lin}_{\delta\delta}(k,z)$, in the LA model predictions for the II and GI terms.
Although it does not provide a fully consistent treatment of non-linear contributions to IA, the NLA model attempts to include the contribution of non-linear structure growth to the tidal field, and it has been shown to provide a better fit to data at quasi-linear scales than the LA model \citep{BK07,SMM15}. 

We also consider a new model, described in \citet{BVS15}, which includes all terms that contribute at next-to-leading order in the tidal alignment scenario, while simultaneously smoothing the tidal field (e.g.\ at the Lagrangian radius of the host halo). The effects of weighting by the source galaxy density can be larger than the correction from the non-linear evolution of dark matter density. This more complete tidal alignment model (denoted the ``CTA model'' below) is described in more detail in Appendix \ref{sec:IA_appendix}.

The left panel of Figure \ref{fig:IA_fid_LA_JAB_noIA} shows cosmological constraints for the
fiducial (NLA),  LA, and CTA models, as well as the case in which IAs are ignored. 
These constraints include marginalization over a free IA amplitude parameter, $A$, with a flat prior over the range [-5,5].
As shown by the values in Table \ref{table:om_s8} and illustrated in Figure \ref{fig:s8om_values}, cosmological parameters are robust to the choice of IA model. The largest departure from the fiducial model happens when IAs are ignored entirely. This decreases the best-fit $S_8$ by
roughly two thirds of the $1\sigma$ uncertainty.
Results for all IA models retain the other choices of our fiducial analysis, including cuts on scale and the choice of cosmological and other nuisance parameters that are marginalised.

The NLA model assumes a particular evolution with redshift, based on the principle that the alignment of galaxy shapes is laid down at some early epoch of galaxy formation and retains that level of alignment afterwards.\footnote{See \citet{KRH+12} and \citet{BVS15} for further discussion of the treatment of non-linear density evolution in the NLA and similar models.}
We can test for more general redshift evolution through the inclusion of a free power-law in $(1+z)$, 
$\eta_{\rm other}$, which we vary within the (flat) prior range [-5,5] and marginalise over, in addition to the IA amplitude free parameter, $A$. Details of these terms and of our IA models are explained in more detail in Appendix \ref{sec:IA_appendix}. 

Our fiducial constraints rely on our ability to constrain the free IA amplitude parameter $A$. We can do this with our standard three-bin tomography because the cosmic shear and IA terms evolve differently with redshift, meaning they contribute with different weight to the observed signal from each bin pair. In the right panel of Figure \ref{fig:IA_fid_LA_JAB_noIA} 
we show constraints on $S_8$ and the IA amplitude, $A$, for our fiducial NLA model with three-bin tomography as well as after marginalising over the redshift power law $\eta_{\rm other}$.
We also show the constraints from an analysis of the fiducial NLA model (no redshift power law) without tomography. 

This figure clearly demonstrates the need for redshift information to constrain the IA contribution. Using three tomographic bins and our fiducial NLA model we obtain a constraint on the IA amplitude which is entirely consistent with $A=1$, although the contours are wide enough that it is also marginally consistent with zero IAs. As soon as the redshift information is reduced, either by using only a single tomographic bin, or by marginalising over an additional power law in redshift, the constraints on the IA amplitude degrade markedly, becoming nearly as broad as our prior range in each case. The constraints on cosmology are also significantly degraded, an effect which is almost entirely due to the degeneracy between the lensing amplitude and the (now largely unconstrained) IA amplitude. The constraints on $S_8$ are considerably stronger if we ignore IAs in the case without tomography.

The use of the free power law in redshift substantially reduces the best-fit value of $S_8$ as well as greatly increasing the errors, as shown in Table \ref{table:om_s8} and Figure \ref{fig:s8om_values}. This is driven by the preference of this model for low values of $\sigma_8$ and $\Omega_m$ when sampling at the negative end of the prior range in $A$. Motivated by astrophysical arguments and observational evidence that red galaxies exhibit radial alignment with overdensities (i.e.\ $A>0$) while blue galaxies are weakly aligned \citep[e.g.][]{SMM15,JMA+11,MBB+11}, we repeat the analysis restricting $A>0$. As expected, imposing this lower bound significantly improves constraints when flexible redshift evolution of IA is allowed (see Table~\ref{table:om_s8} and Figure~\ref{fig:s8om_values}). While allowing for mildly negative $A$ within the tidal alignment paradigm may partially account for potential non-zero alignments of blue and mixed-population source galaxies, a more sophisticated treatment (e.g.\ including ``tidal torquing'' of spiral galaxy angular momenta) should be included in the analysis of future weak lensing measurements with increased statistical power.

\subsection{Matter power spectrum uncertainty}
\label{subsec:nonlin}

\begin{figure}
\includegraphics[width=\columnwidth]{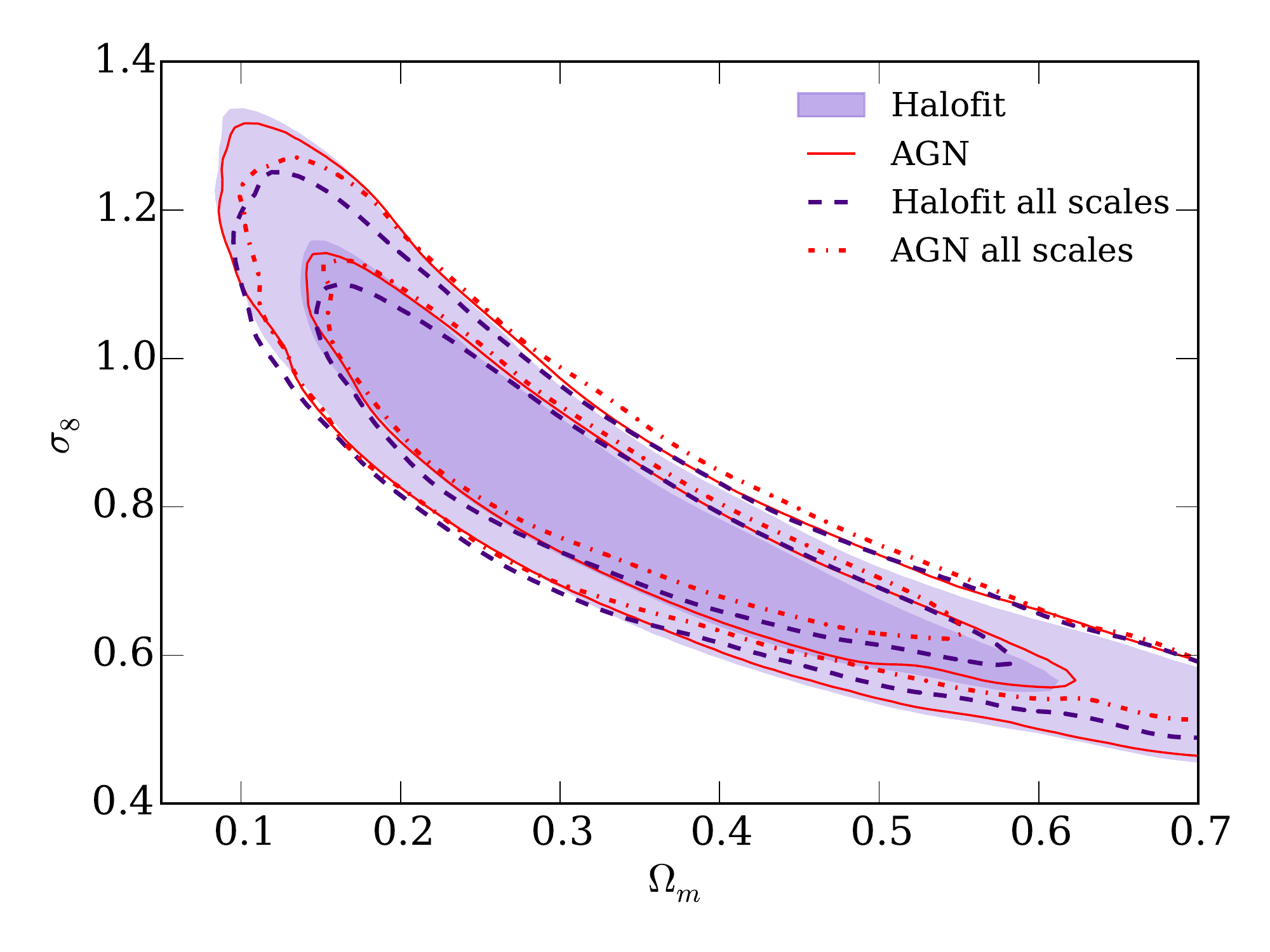}
\caption{
The effect of AGN feedback on cosmological constraints. 
The purple shaded region and the red solid lines use the our fiducial matter power spectrum (\halofit) and the OWLS AGN model respecitvely.
Blue dashed and red dot-dashed lines use a more aggressive data vector, using scales down to 2 arcmin in $\xip$ and $\xim$, again with the fiducial matter power spectrum (\halofit) and the OWLS AGN model respectively.}
\label{fig:baryons}
\end{figure}

Along with IAs, 
the main theoretical uncertainty 
in cosmic shear is 
the prediction of how matter clusters on non-linear scales. For the scales which our measurements are most sensitive to, we require simulations to predict the 
matter power spectrum $\pofk$.

Under the assumption that only gravity affects the matter clustering, \citet{heitmann2014} used the Coyote Universe simulations to achieve an accuracy in $\pofk$ of 1\% at $k\sim1 \mathrm{Mpc}^{-1}$ and $z<1$, and 5\% for $k<10 \mathrm{Mpc}^{-1}$ and $z<4$, a level of error which would have little impact on the results described in this paper. For use in parameter estimation, they released the emulator code FrankenEmu to predict the matter power spectrum given a set of input cosmological parameters. For the range of scales we used in this work, we find very close agreement between \halofit~and FrankenEmu, as demonstrated in Figure \ref{fig:xi_measurements}. We can therefore use \halofit~for our fiducial analysis. However, these codes are based on gravity-only (often referred to as `dark matter-only') simulations which do not tell the whole story. Baryonic effects on the power spectrum due to active galactic nuclei (AGN), gas cooling, and supernovae could be of order 10\% at $k=1\ \rm{Mpc}^{-1}$ \citep{vandalen11}. To predict these effects accurately requires hydrodynamic simulations, which are not only more computationally expensive, 
but are also sensitive to poorly understood physical processes operating well below the resolution scales of the simulations. The effect of baryonic feedback on the matter power spectrum at small scales is therefore sensitive to `sub-grid' physics. 
See \citet{jing06} and \citet{rudd08} for early applications of hydrodynamic simulations in this context, and \citet{vogelsberger14} and \citet{schaye15} for the current state of the art.

As discussed in Section \ref{subsec:scales}, 
in this paper we reduce the impact of non-linearities and baryonic feedback by excluding small angular scales from our data vector. To get an idea of the magnitude of these effects, we have analysed the power spectra from \citet{vandalen11}
which are based on the OWLS simulations (a suite of hydrodynamic simulations which include various different baryonic scenarios).
For a given baryonic scenario, we follow \citet{kitching14} and \citet{maccrann2015} by modulating our fiducial matter power spectrum $P(k,z)$ (from {\sc Camb} and \halofit) as follows:
\be
P(k,z) \rightarrow \frac{P_{\rm{baryonic}}(k,z)}{P_{\rm{DMONLY}}}P(k,z)
\label{eq:owls}
\ee
where $P_{\rm{baryonic}}(k,z)$ is the OWLS power spectrum for a particular baryonic scenario, and $P_{\rm{DMONLY}}$ is the power spectrum from the OWLS `DMONLY' simulation, which does not include any baryonic effects. We assume this somewhat ad-hoc approach of applying a cosmology-independent correction to the cosmology-dependent fiducial matter power spectrum is sufficient for estimating the order of the biases in our constraints  expected from ignoring baryonic effects.
\citet{mccarthy11} find that of the OWLS models, the AGN model best matches observed properties of galaxy groups, both in the X-ray and the optical. 
Furthermore \citet{sembolini11}, \citet{zentner12}, and \citet{eifler14} examine the impact of various baryonic scenarios on cosmic shear measurements, and find that the AGN model causes the largest deviation from the pure dark matter scenario, substantially suppressing power on small and medium scales. 
Of the hydrodynamic simulations we have investigated, the OWLS AGN feedback model is the only one that affects our results significantly, and so we focus on this model here.

Figure \ref{fig:baryons} shows the constraints resulting when performing the modulation above on the 
matter power spectrum, using the AGN model as the baryonic prescription. The purple shaded region and red solid lines, which have small scales removed as described in Section \ref{subsec:scales}, 
are very similar to each other, 
indicating that our choice of scale cuts is conservative, and suggesting that our results are robust to baryonic effects on the power spectrum. The blue dashed and red dot-dashed lines show the constraints when not cutting any small scales from our data vector (i.e. using down to 2 arcminutes in both $\xip$ and $\xim$). Here more of a shift in the constraints is apparent. 
This is quantified in Table \ref{table:om_s8} and illustrated in Figure \ref{fig:s8om_values}. When we use all scales down to 2 arcminutes, the inclusion of the AGN model causes an increase in $S_8$ of 20\% of our error bar (compare the ``Without small-scale cuts'' line 
in Table \ref{table:om_s8} with the ``OWLS AGN $P(k)$ w/o small-scale cuts'' line). 
However, with our fiducial cuts to small scales the increase is only 13\% of our error bar (compare the ``OWLS AGN $P(k)$'' line in Table \ref{table:om_s8} with the Fiducial line). We note that although the contours in Figure \ref{fig:baryons} do appear to tighten slightly along the degeneracy direction when including small scales, the errorbar on $S_8$ increases slightly. This could be due to the theoretical model being a poor fit at small scales, or the noisiness of the covariance matrix.

To take advantage of the small scale information in future weak lensing analyses, more advanced methods of accounting for baryonic effects will be required. \citet{eifler14} propose a PCA marginalisation approach that uses information from a range of hydrodynamic simulations, while \citet{zentner13} and \citet{mead15} propose modified halo model approaches to modelling baryonic effects. Even with more advanced approaches to baryonic effects, future cosmic shear studies will have to overcome other systematics that affect small angular scales, such as the shape measurement selection biases explored in \citet{hartlap11}.

\section{Other data}\label{sec:combination}

In this Section we compare the DES SV cosmic shear constraints with other recent cosmological data. We first compare our results to those from CFHTLenS.
We then compare and combine with the Cosmic Microwave Background (CMB) constraints from Planck (Planck XIII 2015), primarily using the TT + lowP dataset throughout (which we refer to simply as ``Planck'' in most figures).  We also compare to another Planck data combination which used high-$\ell$ TT, TE and EE data and low-$\ell$ 
polarisation 
data.

Planck also measured gravitational lensing of the CMB, which probes a very similar quantity to cosmic shear, but weighted to higher redshifts ($z\sim2$);
we refer to this as ``Planck lensing" when comparing constraints. 
We discuss additional datasets and present constraints on the dark energy equation of state.
See \citet{planck2015cosmo} and \citet{liddlelahav14} for a broad review of current cosmological constraints.

\subsection{Comparisons}

A comparison of DES SV constraints 
to those from other observables is shown in Figure \ref{fig:comparison}.  The observables shown are described below. Constraints on $S_8$ from these comparisons are also shown in Table \ref{table:om_s8} and Figure \ref{fig:s8om_values}.

\label{sec:comparisons}
\subsubsection{Other lensing data}\label{subsec:otherlensing}

\begin{figure}
\includegraphics[width=\columnwidth]{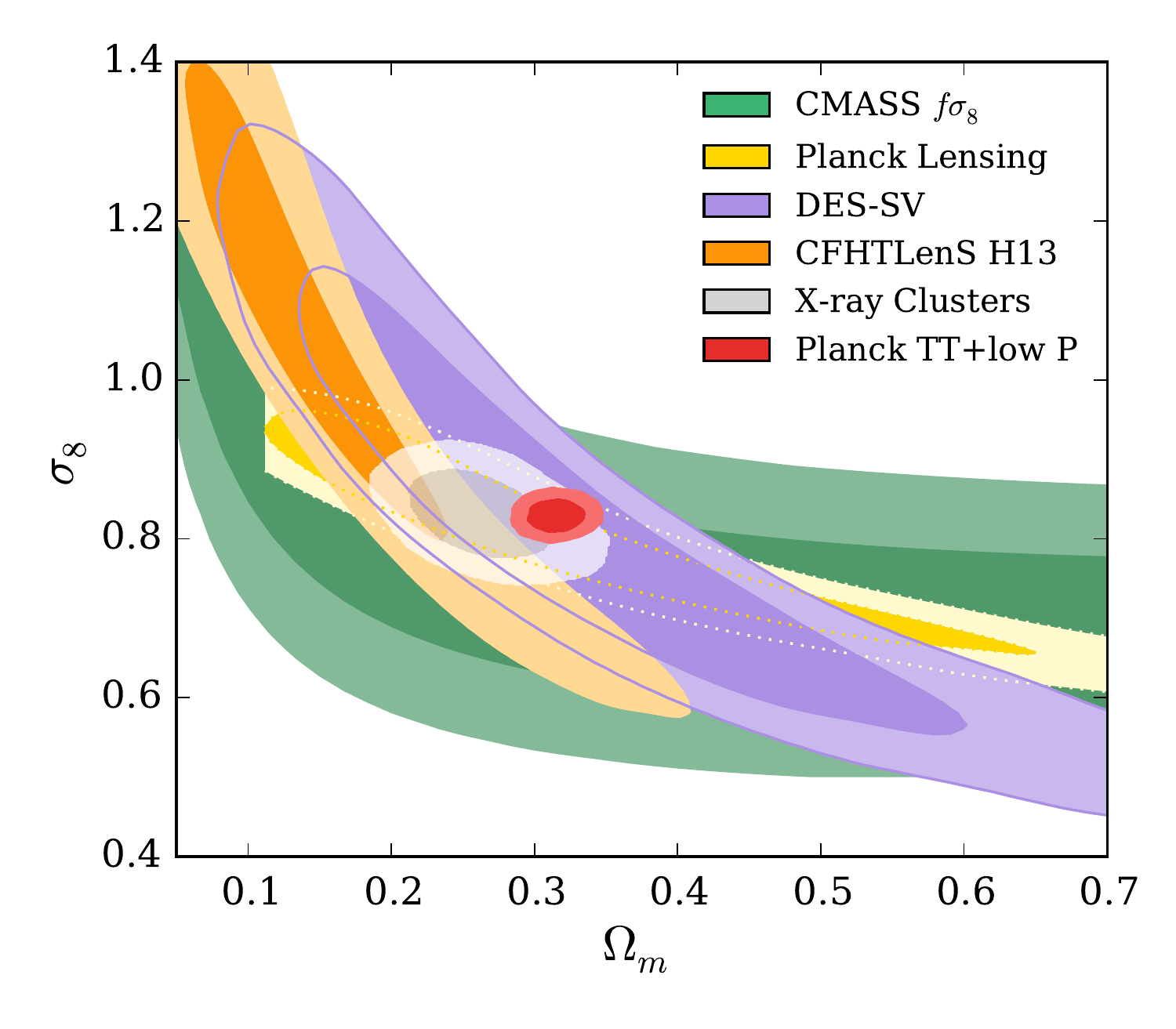}
\caption{
Joint constraints from a selection of recent datasets on the total matter density $\Omega_m$ and amplitude of matter fluctuations $\sigma_8$.  From highest layer to lowest layer:  Planck TT + lowP (red);  X-ray cluster mass counts (\citealt{mantz15}, white/grey shading); DES SV (purple); CFHTLenS (\citetalias{heymans13}, orange);  Planck CMB lensing (yellow); CMASS $f\sig$ (\citealt{Chuang13}, green). 
}
\label{fig:comparison}
\end{figure}

CFHTLenS remains the most powerful current cosmic shear survey, with 154 square degrees of data in the $u$, $g$, $r$, $i$, and $z$ bands.
Table \ref{table:om_s8} 
summarises the constraints from the non-tomographic analysis of \citetalias{kilbinger13} and the tomographic analysis of \citetalias{heymans13} 
that we have computed using the same parameter estimation pipeline as the DES SV data 
(starting from the published correlation functions and covariance matrices).

We investigate the effect of the scale cuts used for the CFHTLenS analysis so that we can make a more fair comparison to DES SV.
In Table \ref{table:om_s8} and Fig \ref{fig:s8om_values}  we show constraints using scale cuts that were used in both \citetalias{heymans13} and \citetalias{kilbinger13} to test the robustness of the results, labelled ``original conservative scales" 
(\citetalias{heymans13} exclude angles $<3'$ for redshift bin combinations involving the lowest two redshift bins from $\xip$, and excluding angles $<30'$ for bin combinations involving the lowest four redshift bins, and angles $<16'$ for bin combinations involving the highest two redshift bins from $\xim$. \citetalias{kilbinger13} exclude angles $<17'$ from $\xip$ and $<53'$ from $\xim$).
Finally, we show the CFHTLenS 
results using minimum scales 
selected using the approach described in Section \ref{subsec:scales}, which we refer to as ``modified conservative scales" in Table \ref{table:om_s8} and Fig \ref{fig:s8om_values}.

We show constraints from~\citetalias{heymans13}, with our scale cuts, on $(\Omega_m, \sigma_8)$ as orange contours in Figure \ref{fig:comparison}.
Our cosmological constraints are consistent with \citetalias{heymans13}, but have a higher amplitude and larger uncertainties. 

The values in Table \ref{table:om_s8} show that our prescription for selecting which scales to use gives similar results to the prescription in \citetalias{heymans13} (compare the ``CFHTLenS (H13) original conservative scales" line to the ``CFHTLenS (H13) modified conservative scales" line). The \citetalias{kilbinger13} results show some sensitivity to switching from using all scales to cutting small scales (possibly because of the apparent lack of power in the large scale points that \citetalias{kilbinger13} used but \citetalias{heymans13} did not), with a lower amplitude preferred when excluding small scales (though see also \citet{kitching14} which prefers higher amplitudes). The uncertainties increase by $\sim 50$\% for  the ``modified conservative scales'' case ($\theta_{min}(\xip)=3.5'$ and $\theta_{min}(\xim)=28'$) compared to using all scales.

The most comparable lines in Table \ref{table:om_s8} show that our tomographic uncertainties are $\sim 20\%$ larger than those from CFHTLenS (compare ``No photoz or shear systematics" with ``CFHTLenS (H13) modified conservative scales")
The main differences between the two datasets are
(i) the DES SV imaging data are shallower and have a larger average PSF than CFHTLenS  
(ii) we are more conservative in our selection of source galaxies (see \citetalias{J15}) 
(iii) we use a larger area of sky (our $139\deg^2$ square degrees instead of 75\% of 154$\deg^2$ $\sim 115\deg^2$; \citet{heymans12}) although our sky area is contiguous instead of four independent patches. 
The upshot of the different depths and galaxy selection are that CFHTLenS has an effective source density of $\sim 11$ per ${\mathrm{arcmin}}^2$ 
while DES SV has an effective density of 6.8 and 4.1 galaxies per ${\mathrm{arcmin}}^2$ for \ngmix~and \imshape~respectively, using the \citetalias{heymans13} definition. 
While the extra redshift resolution in the 6-redshift-bin \citetalias{heymans13} analysis may contribute to their better constraining power (particularly on intrinsic alignments), we expect the main contribution comes from their increased number density of galaxies.
Given the size of our errors, we do not yet have the constraining power required to resolve the apparent discrepancy in the $\Omega_M$ vs $\sigma_8$ plane between CFHTLenS and Planck \citep{maccrann2015,leistedt2014,battye2014}, 
and 
we are consistent with both.

We also show in Table \ref{table:om_s8} and Figure \ref{fig:s8om_values} the result of combining CFHTLenS and DES SV constraints together, which is is straightforward since the surveys do not overlap on the sky. As expected, the joint constraints lie between the two individual constraints. Although judging agreement between multi-dimensional contours is non-trivial, by the simple metric of difference in best-fit $S_8$ divided by the lensing error bar on $S_8$, the tension between CFHTLenS and Planck is somewhat reduced by combining CFHTLenS with DES SV.

Our constraints are also in good agreement with those from Planck lensing \citep{planck2015lensing}, which are shown as yellow contours in Figure \ref{fig:comparison}. The Planck lensing measurement constrains a flatter degeneracy direction in $(\Omega_m, \sigma_8)$ because it probes higher redshifts than galaxy lensing, as discussed in \citet{planck2015lensing} , \citet{pan14}, 
and \citet{jain97}. This means that the constraints it imposes on $\sigma_8 (\Omega_m/0.3)^{0.5}$ are rather weak, as shown in Table \ref{table:om_s8} and Figure \ref{fig:s8om_values}, but the constraints with the best fitting combination $\sigma_8 (\Omega_m/0.3)^{0.24}$ are much stronger (also shown in Table \ref{table:om_s8}).

\subsubsection{Non-lensing data}

Figure \ref{fig:comparison} clearly shows that 
DES SV 
agrees well with Planck on marginalising into
the $\sig-\om$ plane in $\Lambda$CDM. We see in Table \ref{table:om_s8} that this is true for both the Planck TT + lowP and the 
TT+TE+EE+lowP 
variant of the Planck data. 
Since the DES-SV constraints show very little constraining power on any of the other \lcdm parameters varied, agreement of the multi-dimensional contours with Planck seems likely. Since submission of this paper, Raveri 2015 used a Bayesian data concordance test to judge agreement between the constraints from different datasets, including Planck and CFHTLenS. They apply ‘ultra-conservative’ cuts to the CFHTLenS data, resulting in much enlarged contours in the $\om$ - $\sig$ plane, which appear to be in agreement with Planck, however their data concordance test still suggests disagreement between the two datasets. A natural question is whether the converse situation is also be possible - where 2d marginalised contours disagree, but a data concordance test will not show tension. It is clear that caution must be exercised when judging agreement based on 2d marginalised contours.

At the time of writing, the Planck 2015 likelihood code has not been released, but chains derived from it are publicly available. 
As we therefore cannot calculate likelihoods for general parameter choices, we must instead combine Planck with DES SV data using importance sampling: each sample in the Planck chain is given an additional weight according to their likelihood under DES SV data.
Since the Planck chains do not, of course, include our nuisance parameters we must also generate a sample of each of those from our prior to append to each Planck sample. In this approach we must also then \emph{not} apply the nuisance parameter priors again when computing our posteriors during sampling, since that would count the prior twice.
As usual in importance sampling for a finite number of samples this procedure is only valid when the distributions are broadly in agreement, as in this case.
Table  \ref{table:om_s8}  shows that the Planck uncertainties on $S_8$ are reduced by 10\% on combining with DES SV, and the central value moves down by about 10\% of the error bar.
This can be compared to the combination of Planck with Planck Lensing, which brings $S_8$ down further and tightens the error bar more. 

Galaxy cluster counts are a long-standing probe of the matter density and the amplitude of fluctuations (see \citet{mantz15} for a recent review). The constraints from the Sunyayev--Zel'dovich effect measured by Planck \citep{plancksz} are at the lower end of the amplitudes allowed by the DES SV cosmic shear constraints and are in some tension with those from the Planck TT+ lowP primordial constraints, depending on the choice of mass calibration used. X-ray cluster counts also rely on a mass calibration to constrain cosmology and tend to fall at the lower end of the normalisation range (see e.g. \citet{vikhlinin09}). Finally, optical and X-ray surveys can use lensing to measure cluster masses and abundances; there are several ongoing analyses in DES to place constraints on the cluster mass calibration. Figure \ref{fig:comparison} includes a constraint in white from an analysis of X-ray clusters with masses calibrated using weak lensing from \citet{mantz15}.
This is clearly in good agreement with the DES SV results presented here.

Spectroscopic large-scale structure measurements with anisotropic clustering, such as the CMASS data presented in \citet{Chuang13}, can be used to constrain the growth rate of fluctuations, and are shown in green in Figure \ref{fig:comparison}. There is a broad region of overlap between that data and DES SV. 

The Planck 2015 data release contains chains that have been importance sampled with 
large scale structure data from 6dFGS, SDSS-MGS and BOSS-LOWZ \citep{beutler2011,ross2014,anderson14}, supernova data from the Joint Likelihood Analysis \citep{betoule14}, and a re-analysis of the \citet{riess11} HST Cepheid data by \citet{efstathiou14}. 
In Table \ref{table:om_s8} and Figure \ref{fig:s8om_values} we we refer to this combination as `ext' and include it in our importance sampling. 
Planck alone measures $\sig(\om/0.3)^{0.5}= 0.850\pm 0.024$, while Planck+ext measures $\sig(\om/0.3)^{0.5}= 0.824\pm 0.013$.

\begin{figure}
\includegraphics[width=\columnwidth]{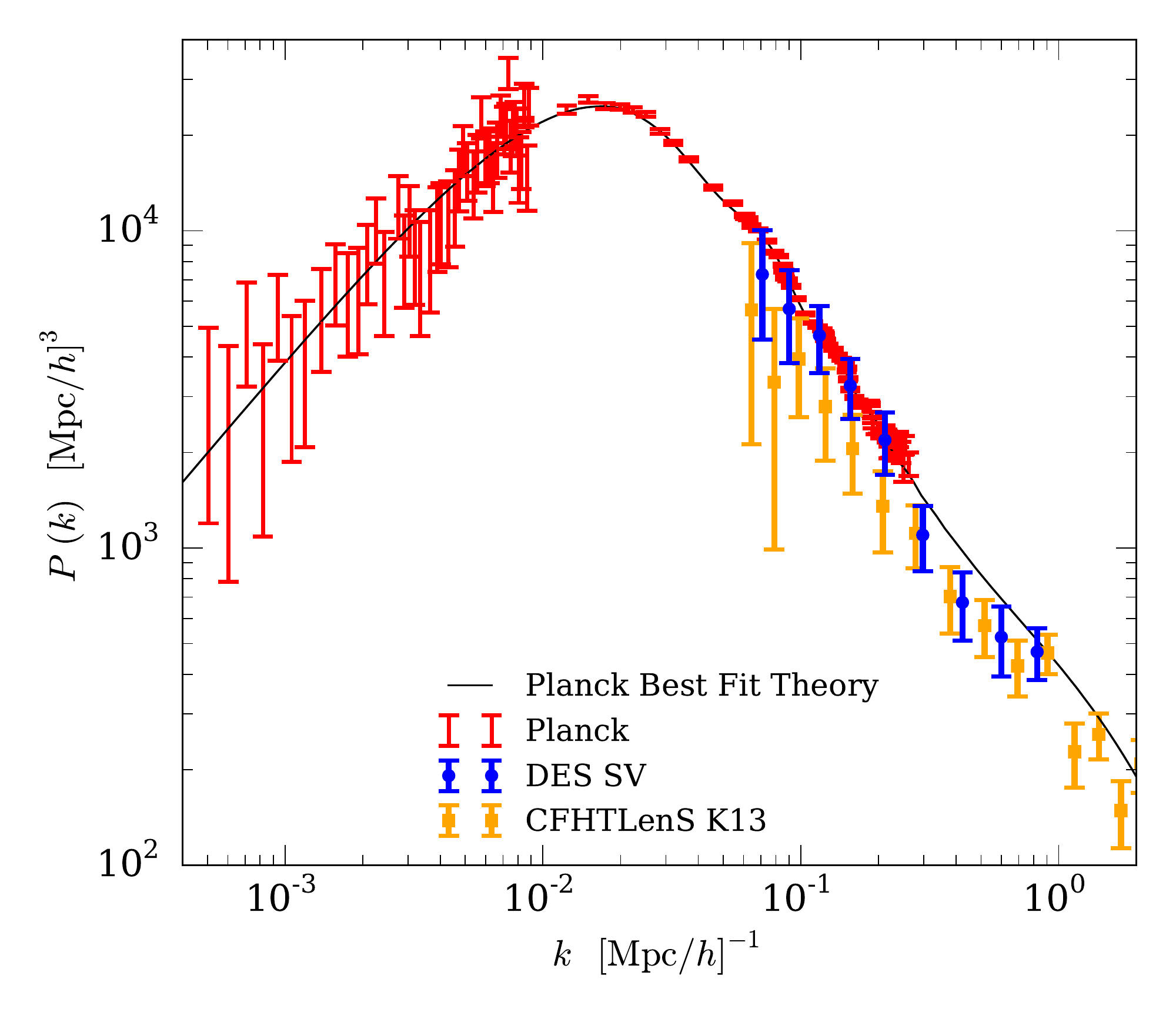}
\caption{
Non-tomographic DES SV (blue circles), CFHTLenS K13 (orange squares) and Planck (red) data points projected onto the matter power spectrum (black line). This projection is cosmology-dependent and assumes the Planck best fit cosmology in $\Lambda$CDM. 
The Planck error bars change size abruptly because the $C_{\ell}$s are binned in larger $\ell$ bins above $\ell=50$.
}
\label{fig:pk}
\end{figure}

Figure \ref{fig:pk} shows the DES SV, CFHTLenS and Planck data points translated onto the matter power spectrum assuming a $\Lambda$CDM cosmology. This uses the method described in \citet{maccrann2015} which follows \citet{tegmark2002} in translating the central $\theta$ and $\ell$ values of the measurements into wavenumber values $k$. The wavenumber of the point is the median of the window function of the $P(k)$ integral used to predict the observable ($\xip$ or $C_{\ell}$). The height of the point is given by the ratio of the observed to predicted observable, multiplied by the theory power spectrum at that wavenumber. 
For simplicity we use the no-tomography results from each of DES SV and CFHTLenS (K13). 
The results are therefore cosmology dependent, and we use the Planck best fit cosmology for the version shown here. 
The CFHTLenS results are below the Planck best fit at almost all scales (see also discussion in \citealt{maccrann2015}). 
The DES results agree relatively well with Planck up to the maximum wavenumber probed by Planck, and then drop towards the CFHTLenS results.

\begin{figure}
\includegraphics[width=\columnwidth]{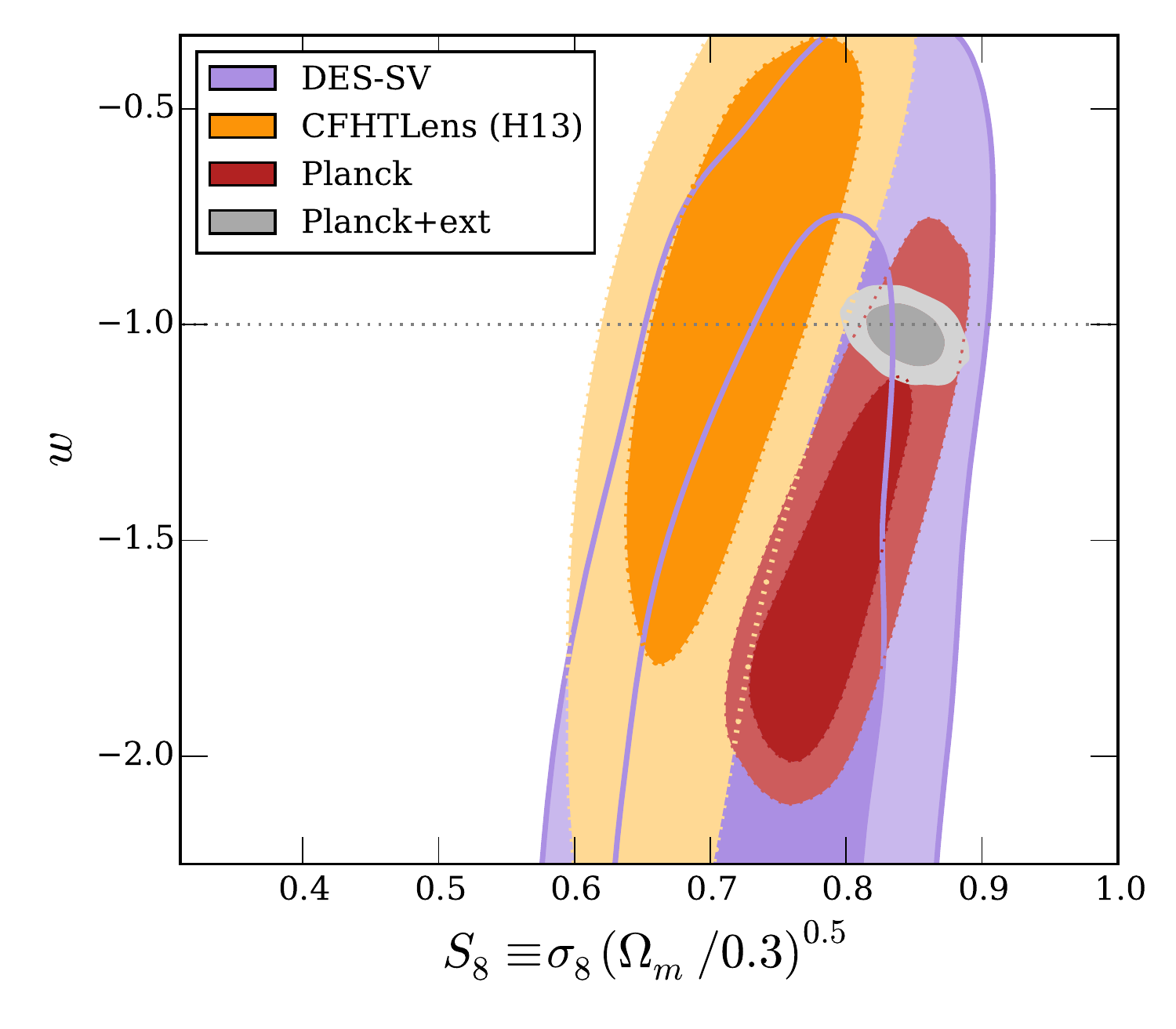}
\caption{
Constraints on the dark energy equation of state $w$ and $S_8\equiv \sigma_8 (\Omega_m/0.3)^{0.5}$, from DES SV (purple), Planck (red), CFHTLenS (orange), and Planck+ext (grey).  DES SV is consistent with Planck at $w=-1$.  The constraints on $S_8$ from DES SV alone are also generally robust to variation in $w$. }
\label{fig:w_scatter}
\end{figure}

\subsection{Dark Energy}

The DES SV data is only 3\% of the total area of the full DES survey, so we do not expect to be able to significantly constrain dark energy with this data. Nonetheless, we have recomputed the fiducial DES SV constraints for the second simplest dark energy model, $w$CDM, which has a free (but constant with redshift) equation of state parameter $w$, in addition to the other cosmological and fiducial nuisance parameters (see Section 3). The purple contours in Figure \ref{fig:w_scatter} show constraints on $w$ versus the main cosmic shear parameter $S_8$; we find DES SV has a slight preference for lower values of $w$, with $w<-0.68$ at 95\% confidence. There is a small positive correlation between $w$ and $S_8$, but our constraints on $S_8$ are generally robust to variation in $w$.

The Planck constraints (the red contours in Figure \ref{fig:w_scatter}) agree well with the DES SV constraints: combining DES SV with Planck gives negligibly different results to Planck alone. This is also the case when combining with the Planck+ext results shown in grey. \citet{planck2015} discuss that while Planck CMB temperature data alone do not strongly constrain w, they do appear to show close to a $2\sigma$ preference for $w<-1$. However, they attribute it partly to a parameter volume effect, and note that the values of other cosmological parameters in much of the $w<-1$ region are ruled out by other datasets (such as those used in the `ext' combination).

 Planck CMB data combined with CFHTLenS also show a preference for $w<-1$ \citep{planck2015}. The CFHTLenS constraints (orange contours) in Figure \ref{fig:w_scatter} show a similar degeneracy direction to the DES SV results, although with a preference for slightly higher values of $w$ and lower $S_8$. The tension between Planck and CFHTLenS in $\Lambda$CDM is visible at $w=-1$, and interestingly, is not fully resolved at any value of $w$ in Figure \ref{fig:w_scatter}. This casts doubt on the validity of combining the two datasets in $w$CDM. 

\section{Conclusions}
\label{sec:conclusions}

We have presented the first constraints on cosmology from the Dark Energy Survey. Using 139 square degrees of Science Verification data we have constrained the matter density of the Universe $\Omega_{\rm m}$ and the amplitude of fluctuations $\sigma_8$, and find that the tightest constraints are placed on the degenerate combination $S_8 \equiv\sigma_8( \Omega_{\rm m}/0.3)^{0.5}$, which we measure to 7\% accuracy to be $S_8=0.81\pm0.06$.

DES SV alone places weak constraints on the dark energy equation of state: $w<-0.68$ (95\%). These do not significantly change constraints  on $w$ compared to Planck alone, and the cosmological constant remains within marginalised DES SV+Planck contours.

The state of the art in cosmic shear, CFHTLenS, gives rise to some tension when compared with the most powerful dataset in cosmology, Planck \citep{planck2015cosmo}. Our constraints are in agreement with both Planck and CFHTLenS results, and we cannot rule either out due to larger uncertainties caused by a smaller effective number density of galaxies and our propagation of uncertainties in the two most significant lensing systematics into our constraints. 

We have investigated the sensitivity of our results to variation in a wide range of aspects of our analysis, and found our fiducial constraints to be remarkably robust.
Our results are stable to switching to our alternative shear catalogue, \imshape, or to any of our alternative photometric redshift catalogues, TPZ, ANNZ2 and BPZ. Nonetheless, to account for any residual systematic error we marginalise over 5\% uncertainties on shear and photometric redshift calibration in each of three redshift bins in our fiducial analysis; this inflates the error bar by 9\%. 

Our results are also robust to the choice of data vector: constraints from Fourier space $C_\ell$  are consistent with those from real space $\xipm(\theta)$. As expected, a 2D analysis is less powerful than one split into redshift bins; the biggest benefit of tomography comes from its constraints on intrinsic alignments.

In the future, DES will be an excellent tool for learning about the nature of IAs. In this current analysis we only aim to show that the details of IA modelling do not affect the cosmological conclusions drawn from the SV dataset.  We investigated four alternatives to our fiducial intrinsic alignment model and found the results to be stable, even when including an additional free parameter adding redshift dependence. 
Similarly, the similarity in parameter constraints when using the NLA and CTA models, as well as the minor shift when compared with the LA and no IA cases, is consistent with the results of \citet{KEB15}, who forecast the effects of IA contamination for each of these models for the full DES survey.

The DES SV results are also robust to astrophysical systematics
in the matter power spectrum predictions. We chose to use only scales where the effect of baryons on the matter power spectrum predictions are expected to be relatively small, 
however, our results are relatively insensitive to the inclusion of small angular scales and to the effects of
baryonic feedback as implemented in the OWLS hydrodynamic simulations. Our fiducial results are shifted by only 14\% of the error bar when the OWLS AGN model is included.

In the analysis of future DES data from Year One and beyond we aim to be more sophisticated in several ways. Greater statistical power will allow us to constrain our astrophysical systematics more precisely, and algorithmic improvements will reduce our nuisance parameter priors.  Forthcoming Dark Energy Survey data will provide 
much more powerful cosmological tests, such as constraints on neutrino masses, modified gravity, and of course dark energy.

\section*{Acknowledgements}

We are grateful for the extraordinary contributions of our CTIO colleagues and the DECam Construction, Commissioning and Science Verification
teams in achieving the excellent instrument and telescope conditions that have made this work possible.  The success of this project also 
relies critically on the expertise and dedication of the DES Data Management group.

We are very grateful to Iain Murray for advice on importance sampling. We thank Catherine Heymans, Martin Kilbinger, Antony Lewis and Adam Moss for helpful discussion. 

This paper is DES paper DES-2015-0076 and FermiLab preprint number FERMILAB-PUB-15-285-AE.

Sheldon is supported by DoE grant DE-AC02-98CH10886.
Gruen was supported by SFB-Transregio 33 `The Dark Universe' by the Deutsche Forschungsgemeinschaft (DFG) and the DFG cluster of excellence `Origin and Structure of the Universe'.
Gangkofner acknowledges the support by the DFG Cluster of Excellence `Origin and Structure of the Universe'.
Jarvis has been supported on this project by NSF grants AST-0812790 and AST-1138729.
Jarvis, Bernstein, and Jain are partially supported by DoE grant DE-SC0007901. 
Melchior was supported by DoE grant DE-FG02-91ER40690.
Plazas was supported by DoE grant DE-AC02-98CH10886 and by JPL, run by Caltech under a contract for NASA. 

Funding for the DES Projects has been provided by the U.S. Department of Energy, the U.S. National Science Foundation, the Ministry of Science and Education of Spain, 
the Science and Technology Facilities Council of the United Kingdom, the Higher Education Funding Council for England, the National Center for Supercomputing 
Applications at the University of Illinois at Urbana-Champaign, the Kavli Institute of Cosmological Physics at the University of Chicago, 
the Center for Cosmology and Astro-Particle Physics at the Ohio State University,
the Mitchell Institute for Fundamental Physics and Astronomy at Texas A\&M University, Financiadora de Estudos e Projetos, 
Funda{\c c}{\~a}o Carlos Chagas Filho de Amparo {\`a} Pesquisa do Estado do Rio de Janeiro, Conselho Nacional de Desenvolvimento Cient{\'i}fico e Tecnol{\'o}gico and 
the Minist{\'e}rio da Ci{\^e}ncia, Tecnologia e Inova{\c c}{\~a}o, the Deutsche Forschungsgemeinschaft and the Collaborating Institutions in the Dark Energy Survey. 
The DES data management system is supported by the National Science Foundation under Grant Number AST-1138766.

The Collaborating Institutions are Argonne National Laboratory, the University of California at Santa Cruz, the University of Cambridge, Centro de Investigaciones En{\'e}rgeticas, 
Medioambientales y Tecnol{\'o}gicas-Madrid, the University of Chicago, University College London, the DES-Brazil Consortium, the University of Edinburgh, 
the Eidgen{\"o}ssische Technische Hochschule (ETH) Z{\"u}rich, 
Fermi National Accelerator Laboratory, the University of Illinois at Urbana-Champaign, the Institut de Ci{\`e}ncies de l'Espai (IEEC/CSIC), 
the Institut de F{\'i}sica d'Altes Energies, Lawrence Berkeley National Laboratory, the Ludwig-Maximilians Universit{\"a}t M{\"u}nchen and the associated Excellence Cluster Universe, 
the University of Michigan, the National Optical Astronomy Observatory, the University of Nottingham, The Ohio State University, the University of Pennsylvania, the University of Portsmouth, 
SLAC National Accelerator Laboratory, Stanford University, the University of Sussex, and Texas A\&M University.

The DES participants from Spanish institutions are partially supported by MINECO under grants AYA2012-39559, ESP2013-48274, FPA2013-47986, and Centro de Excelencia Severo Ochoa SEV-2012-0234 and SEV-2012-0249.
Research leading to these results has received funding from the European Research Council under the European Union Seventh Framework Programme (FP7/2007-2013) including ERC grant agreements 
 240672, 291329, and 306478.
 
 This paper has gone through internal review by the DES collaboration.

\bibliographystyle{mnras}
\bibliography{refs}

\appendix
\section{Intrinsic Alignment Models}
\label{sec:IA_appendix}

Here we briefly describe our fiducial, NLA, model of intrinsic alignments (IAs), as well as the other models we compare against in Section \ref{subsec:IAs}.

The observed 
cosmic shear power spectrum is the sum of the effect due to gravitational lensing, GG, the IA auto-correlation, II, and the gravitational-intrinsic cross-terms:
\begin{equation}
C^{ij}_{\rm obs}(\ell) = C^{ij}_{\rm GG}(\ell) + C^{ij}_{\rm GI}(\ell) + C^{ij}_{\rm IG}(\ell) + C^{ij}_{\rm II}(\ell). 
\end{equation}
When we quote results for ``No IAs'' we are simply ignoring the three IA terms on the right hand side of this equation.

Each of these contributions can be written as integrals over appropriate window functions and power spectra,
\begin{eqnarray}
C^{ij}_{\rm GG}(\ell) &=& \int_{0}^{\chi_{\rm hor}} \frac{dz}{z^{2}} g^{i}(z) g^{j}(z) P_{\delta\delta}(k,z), \\
C^{ij}_{\rm II}(\ell) &=& \int_{0}^{\chi_{\rm hor}} \frac{dz}{z^{2}} n_{i}(z) n_{j}(z) P_{\rm{II}}(k,z), \\
C^{ij}_{\rm GI}(\ell) &=& \int_{0}^{\chi_{\rm hor}} \frac{dz}{z^{2}} g^{i}(z) n_{j}(z) P_{\rm{GI}}(k,z),
\end{eqnarray}
where $g^i(z)$ is the lensing efficiency function, $n_{i}(z)$ is the redshift distribution of the galaxies in tomographic bin $i$ and we have assumed the Limber approximation. The details of any chosen IA model are encoded in the auto- and cross-power spectra, $P_{\rm{II}}$ and $P_{\rm{GI}}$.

Within the tidal alignment paradigm of IAs (see \citet{TI15,JCK+15,KCJ+15,KBH+15} for general reviews of IAs), the leading-order correlations define the linear alignment (LA) model \citep{HS10}. In the LA model predictions for the II and GI terms give
\be
P_{\rm{II}}(k,z) = F^2(z)P_{\delta\delta}(k,z) ,\;\;   P_{\rm{GI}}(k,z) = F(z)P_{\delta\delta}(k,z), \label{eqn:NLA}
\ee
\noindent
where
\be
F(z) = -AC_1\rho_{\textrm{crit}}\frac{\om}{D(z)} .
\ee
\noindent
$\rho_{\rm crit}$ is the critical density at $z=0$, $C_1 = 5 \times 10^{-14}h^{-2}M_{\odot}^{-1}{\rm Mpc}^3$ is a normalisation amplitude \citep{HMR+01,BTB+03,BK07}, and $A$, the dimensionless amplitude, is the single free parameter. $D(z)$ is the growth function. In the case where redshift dependence for IA is included, the amplitude is
\begin{equation}
F(z,\eta_{\rm other}) = -AC_1\rho_{\textrm{crit}}\frac{\om}{D(z)} \left( \frac{1+z}{1+z_{0}} \right)^{\eta_{\rm other}}.
\end{equation}

In the LA alignment paradigm galaxy intrinsic alignments are sourced at the epoch of galaxy formation and do not undergo subsequent evolution, as such they are unaffected by non-linear clustering at late times, and the $P_{\delta\delta}(k,z)$ that enter equation \ref{eqn:NLA} are linear matter power spectra. Our fiducial model, the non-linear alignment (NLA) model, simply replaces the linear power spectra with their non-linear equivalents, $P^{\rm nl}_{\delta\delta}$, wherever they occur, increasing the power of IAs on small scales. This simple ansatz has no physical motivation under the LA paradigm, but it has been shown to agree better with data \citep{BK07,SMM15}. The non-linear power spectra are calculated using the \citet{takahashi2012} version of the \halofit~formalism \citep{SPJ+03}.

We also consider a model called the complete tidal alignment (CTA) model \citep{BVS15}. This model includes all terms that contribute at next-to-leading order in the tidal alignment scenario, while also smoothing the tidal field. The equivalent II and GI terms 
\bea
P_{GI}(k,z) &=& \!\!F_{\rm CTA}(z)\left[P_{\rm NL}(k,z) + \frac{58}{105}b_1\sigma_S^2 P_{\rm lin} + b_1 P_{0|0\mathcal{E}}\right]  \, ,\nonumber \\
P_{II}(k,z)&=& \!\!F_{\rm CTA}^2(z)[P_{\rm NL}(k,z) + \frac{116}{105}b_1\sigma_S^2 P_{\rm lin} \nonumber \\
&& + 2b_1 P_{0|0\mathcal{E}} + b_1^2 P_{0\mathcal{E}|0\mathcal{E}}]\, ,
\eea
where $b_1$ is the linear bias of the source sample (approximated to be $b_1=1$ for our sample), $\sigma_{S}^2$ is the variance of the density field, smoothed in Fourier space at a comoving scale of $k=1~h^{-1}{\rm Mpc}$, corresponding to roughly the Lagrangian radius of a dark matter halo. $P_{0|0\mathcal{E}}$ and $P_{0\mathcal{E}|0\mathcal{E}}$ are $\mathcal{O}(P_{\rm lin}^2)$ terms that arise from weighting the intrinsic shape field by the source density.  
The amplitude of the CTA model is given by 
\be
F_{\rm CTA} = -AC_1\rho_{\textrm{crit}}\om (1+z) \left(1+\frac{58}{105}b_1 \sigma_S^2\right)^{-1}.
\ee

\label{lastpage}

\end{document}